\documentclass{article}
\usepackage[T1]{fontenc}
 
\usepackage{graphicx} 
\usepackage{mathrsfs}
\usepackage{amsmath,amsfonts,amssymb,amsthm}
\usepackage{stmaryrd}
\usepackage{xcolor}
\usepackage{dsfont}
\usepackage{hyphenat}
\usepackage{tikz}
\usetikzlibrary {arrows.meta,math,graphs,calc} 
\usepackage{forest} 
\usepackage[intoc]{nomencl}
\usepackage{xifthen}
\usepackage[backend=biber,style=alphabetic]{biblatex} 
\usepackage{multirow}
\usepackage{hhline}
\usepackage{accents}
\input macros
\title{Structure and statistical organization of\\the stationary state of the Oslo model}
\author{Valentin Lallemant$^1$ and Vincent Rossetto$^1$\\
$^1$\small{\textit{Université Grenoble Alpes, CNRS, LPMMC, 38000 Grenoble, France}}}
\date{August 2025}

\newcommand\CORR[2]{\if\relax\detokenize{#1}\relax\else{\color{red}(#1)}\;\fi{\color{blue}#2}}

\begin{document}

\maketitle
\begin{abstract}
In most driven-dissipative sandpile models, the dynamics of the system reaches a critical stationary state.
This state displays organization features such as a power-law avalanche spectrum and hyperuniformity,
but these features often emerge without a clear path from the microscopic evolution rules. 
Only in a few cases is there an available description of the stationary state, in other 
sandpile models the question is open.
In this article, we present our result on the stationary state of the Oslo model, a driven-dissipative sandpile 
model with intrinsic randomness. In order to do so, we use different representations of the system
configurations and of the dynamical process. Moving back and forth between these representations allows to 
identify invariant quantities for each configurations. 
Moreover, we obtain the detailed statistical description of the stationary state 
by considering all paths leading to a given configuration at once, and by summing their contributions
under the constraint specified by the invariants. As a result, we find that the configurations of the
stationary state are structured into a small number of equivalence classes, and that their
statistical weights are related to the counting of colored diagrams respecting a small set of rules.
\end{abstract}

\section{Motivations and context}
The ubiquity of scale invariant phenomena in out of equilibrium systems is a recurrent theme in statistical physics. 
Almost 40 years ago, Self-Organised Criticality (SOC) has been introduced \cite{bak_self-organized_1987} as an attempt
to explain such spontaneous emergence of critical features in out-of-equilibrium settings. 
It has been successful in inspiring a great varieties of studies across the scientific fields: 
biology, neuroscience, astrophysics, seismology, economy and many more.
Dedicated reviews can be found on each subject (e.g. \cite{dhar_theoretical_2006}, 
\cite{aschwanden_25_2016}, \cite{bouchaud_self-organized_2024} respectively dealing with toy models in physics,
astrophysics and economy fields), and the book of Bak \cite{bak_how_1996}
might give a nice first overview on the wide applicability of the concept.
So to say, SOC applies to a wide class of systems and can be qualified, without much doubts, as a universal 
phenomenon.

Along the way to study SOC,
many models have been introduced, of which the sandpile models form the most prominent class.
However, despite convincing numerical studies along the years, a global understanding,
at the level of mathematical rigor, of these models remains
partial. Indeed, the exact derivation of the model's observables follows from case-by-case considerations
rather than as a universal feature. 

A better understanding, from analytical considerations, arises in two families of models : directed models 
--- such as the Totally Asymetric Oslo model (TAOM)
\cite{pruessner_exact_2004} and directed versions of the BTW model \cite{dhar_exactly_1989} --- and the BTW model itself
(see \cite{jarai_sandpile_2018} for a cover of some of the main results on the subject). 
In both cases, integrability originates from mappings and sometimes also
from an underlying group structure.
The directed models are equivalent to random walks, whereas, for the BTW, a large amount of results
were obtained using a mapping between the stationary state and a problem of spanning tree enumeration on a lattice.
In both cases, static and dynamical quantities were rigorously evaluated.

In general, sandpile models incorporate stochasticity which can be implemented in two manners:
either externally, through the injection and/or the dissipation mechanism, or internally, along the dynamics.
This second scenario has definitely proven to be the hardest to investigate, 
as evidenced by the small amount of exact results for the Activated Random Walk (ARW), Oslo or Manna models. 
One of the main obstruction is the absence of an exact formula for the stationary states of these models, 
preventing access to many local and global observables necessary to get a consistent picture of SOC in sandpile models.
Gaining insight into these types of model faces two different challenges.
First, from a physical point a view, systems
with intrinsic stochasticity take into account a local variability which is relevant in many real
systems involving mesoscopic size agents. 
Second, the analytical barrier in these models 
(absence of a group structure and of mappings with other known problems) asks for new methods of studies.
They are of interest in their own and might well applied to many more studies afterwards.
In this context, the Oslo model (OM) is one of the easiest model to work with, notably because of its dimensionality 
($1$D in space), and the limited number of rules needed to describe its evolution.
This article is dedicated to give an explicit description of its stationary state in a driven-dissipative setup. 
It can be viewed as a continuation of the work of Dhar \cite{dhar_steady_2004} that has, 
to the extent of our knowledge, not been addressed up to now.
Our work strongly relies upon its main results.  

Before diving into formal definitions, we present briefly the genesis of the model.
First of all, the Oslo model is deeply connected to the first experimental tests of self-organised criticality.
Indeed, in \cite{frette_sandpile_1993} and \cite{frette_avalanche_1996}, theoretical and experimental works were
conducted so as to test the assumptions made in \cite{bak_self-organized_1987}.
The Oslo sandpile model emerged in this context.
Its modern form was formalised in \cite{christensen_tracer_1996},
and is historically seen as a refinement of the model introduced in \cite{frette_sandpile_1993}.
Ever since, the model has been studied through many approaches among which analysis
\cite{dhar_steady_2004}, field theory \cite{pruessner_oslo_2003},
symbolic computation \cite{corral_calculation_2004} and computer simulations \cite{grassberger_oslo_2016}.
Our investigations have been guided by numerical tools but our results are analytical.

The mathematical structure of the Oslo model presents difficulties similar to those of other stochastic sandpiles models, 
such as the ARW. These difficulties are addressed here by using different representations of the system, 
each of which presents
specific advantages depending on the properties we want to prove.
The abstract representations have helpful graphical counterparts
that we encourage the reader to use ; they should facilitate the understanding of the technical aspects of the article.

In Section \ref{sect:definitions}, we define the 
model and introduce along the way a new representation of the configurations, that we call the $g$-representation. 
This representation will become handy as it highlights at a glance some conservation
laws of the problem and some of its dynamical properties.  
In Section \ref{sect:def stab}, we introduce notations and objects to describe the dynamical evolution 
in the sandpile largely inspired by the literature on the ARW model.
In Section \ref{sect:classes recur}, we derive the first results of the paper.
We characterize many invariant quantities and find a natural partition of the configurations which allows 
us to factorize the stationary state expression.
In Section \ref{sect:count paths}, we use a mapping between the expression of probabilities in the stationary
state and a coloring problem.
We conclude adding up all the main results and build an explicit formula for the Oslo stationary state,
and open the discussion with some conjectures and direction for future works.

\section{Definition of the model}\label{sect:definitions}
Several representations of the Oslo model have been proposed, each of them shedding light on
different
aspects of its configurations and dynamics. 
Many of the reasoning presented here take advantage of the
existence of these multitude of representations. 

\subsection{Stochastic dynamics}
The Oslo model is a sandpile model originally introduced
\cite{christensen_tracer_1996} to describe rice pile
experiments \cite{frette_avalanche_1996}.
This is a discrete model, defined in one dimension, on a length~$L\in\N$. 
For each $x\in\N$ such that $1\leq x\leq L$, the site~$x$ contains 
$h(x)\in\N$ grains. 
Grains can move from a site to the next site to their right. 
Such moves are called \emph{topplings} and are determined by the 
height difference (also called \emph{slope})
\begin{equation}
z(x) = h(x)-h(x+1).
\label{eq:z}
\end{equation}
Topplings happen depending on the value of a stochastic slope threshold $\zc(x)\in\{1,\,2\}$ 
defined at each site: when $z(x)>\zc(x)$, a toppling happens at~$x$,
after which the value of~$\zc(x)$ is reset to 2 with probability $p$ and 1 with 
probability~$1-p=q$.

Because of this toppling condition, it is customary to represent the 
configuration of the system using~$z$ instead of $h$, by listing the values
for all sites in a ``ket'' like $\zcfg{221}$, that represents a $L=3$ system
with slopes~$2$, $2$ and $1$ from $x=1$, $2$ and $3$ respectively.
Only configurations where~$z$ takes non-negative values are studied.

\begin{definition}[Stable configuration]
We call \emph{stable} any configuration where $z$ only takes the values $0$, $1$ or $2$ ; this
corresponds to configurations for which there is a threshold function~$\zc$ satisfying
$z(x)\leq \zc(x)$ for all $x$. The set of all stable
configurations is denoted by~$\stable$. 
\end{definition}

Since all movements are performed to the right, the site $x=0$ never gets grains 
and acts as a wall. 
A toppling at site~$L$ permanently removes a grain located at site $L$ from the system,
making site~$L+1$ acting as a dissipative sink. This translates into the
convention $h(L+1)=0$ at all times. 

Following Dhar~\cite{dhar_steady_2004}, we consider unstable configurations, that are those 
where at least one site could be modified depending on the relative values of~$z$ and~$\zc$, 
but we describe undetermination using \emph{waiting units} at these sites. 
We denote the number of waiting units at site~$x$ as $w(x)$, and we also use
Dhar's operator notation $a_x$ to denote a waiting unit at~$x$.

\begin{definition}[Generalized configuration]
A generalized configuration is a stable configuration, an element of~$\stable$,
and a multiset (a set in which an element can appear many times) of waiting units. 
We denoted the set of generalized configurations by~$\general$. 
\end{definition}
An example of a generalized configuration is $a_1\zcfg{221}$ 
and we provide in Figure \ref{fig:z generalized} another example using the $z$-representation graphical form.
With these definitions, as soon as a generalized configuration does not contain any 
waiting units, it is stable \cite{dhar_steady_2004}. 

We view the stabilization of the system as a sequence of activations, one
waiting unit at a time, where the activation of a waiting unit determines the next configuration.
This process goes on as long as there are units waiting to be activated, and always ends
at a stable configuration.
When a waiting unit is activated, we say that an \emph{instruction} is performed on the
system, emphasizing the analogy between the stabilization process and a computation.
Table~\ref{tab:cases} below presents the instructions performed by a waiting $z$-unit
at site~$x$ (wherever $w(x)\geq1$) depending on the value of~$z(x)$. Performing an 
instruction modifies $z(x)$ and the value of $w$ at up to three sites. 
This presentation of the Oslo model is equivalent to the original, but it does not use 
the threshold $\zc$ : the stochastic choices, between instructions~\instr{p} 
and \instr{q}, are made on the fly. 

\begin{table}[ht!]
\centering
\caption{\label{tab:cases}The different possible local instructions, in the bulk,
along the stabilization procedure.
A waiting unit $a_x$ is identified in a general configuration, $a_xc$, where $c\in\general$. 
$c^{x}$ is the generalized configuration after the evolution at~$x$, and $z$ and $z^x$
the $z$-representations of $c$ and $c^x$ respectively ; $z$ and $z^x$ are identical, except
at site~$x$. 
(*) For $x=1$ the toppling instructions execute $a_1c\to a_2c^x$
and for $x=L$, they execute $a_Lc\to a_{L-1}a_Lc^x$.
The Figure \ref{fig:rules in z} is a graphical translation of the instructions in $z$-representation.}
\begin{tabular}{|c|c|c|c|l|l|}
\hline
$z(x)$ & probability & $z^x(x)$ & instruction & step & type \\
\hline
\hline
$z(x)=0$ & 1 & $z^x(x)=1$ & \instr{s} & \multirow{2}{*}{$a_xc \to c^x$} 
  & \multirow{2}{*}{settling}\\
\hhline{----~~}
\multirow{2}{*}{$z(x)=1$} & $p$ & $z^x(x)=2$ & \instr{p} & & \\
\hhline{~-----}
& $q=1-p$ & $z^x(x)=0$ & \instr{q} & \multirow{2}{*}{$a_xc \to a_{x-1}a_{x+1}c^x$} &  \multirow{2}{*}{toppling$^*$}\\
\hhline{----~~}
$z(x)=2$ & 1 & $z^x(x)=1$ & \instr{t} & &\\
\hline
\end{tabular}
\end{table}

Finally, we consider the Oslo model in its driven-dissipative setup.
The drive is realized by the addition of one waiting particle at a time at $x=1$.
Between two injections, the system is stabilized, which means that 
topplings are performed at an infinitely larger rate than injection.
The source of stochasticity is therefore restricted
to the random choice between instructions \instr{p} and \instr{q}. 
We call \emph{avalanche} the sequence of unstable configurations starting with the
addition of a particle at $x=1$ and ending with a stable configuration.
Along an avalanche, the number of unstable sites unpredictably fluctuates until it 
reaches zero. 
It was shown by Dhar that the order in which unstable sites topple during an avalanche 
does not affect the statistics of the final stable configurations \cite{dhar_steady_2004},
a property called Abelian symmetry.

\subsection{Representations of generalized configurations}
The $h$-representation, that counts the number of grains on a site,
and the $z$-representation, that we use to define the dynamics, 
contain the same informations for stable configurations. 
In this Section, we extend the $z$-representation for generalized configurations
and we introduce their graphical counterparts. 
We also introduce a third representation, that we will extensively 
use in this article.

\begin{figure}[ht!]
\centering
\begin{tikzpicture}
    \begin{scope}
        \reprzg[black]{1/1,0/0,1/2,1/0,1/1}
        \node[rectangle,draw] at (1.25,-0.5) {$z$};
    \end{scope}
    \begin{scope}[xshift=5cm]
        \reprgg{2/1,2/0,1/2,1/0,0/1}
        \node[rectangle,draw] at (1.25,-0.5) {$g$};
    \end{scope}
\end{tikzpicture}
\caption[Graphical $z$-representation]
{\label{fig:z generalized}
Graphical $z$- and $g$-representations of the generalized configuration 
$c=a_1a_3^2a_5\zcfg{10111}\in\general$ ($c=\zcfg{\w10\ww11\w1}$).
The particles of the stable configuration~$\zcfg{10111}$ are black circles
those of the $g$-representation are white squares.
Waiting units are gray in both representations.} 
\end{figure}

\begin{figure}[ht!]
\centering
\begin{tikzpicture}
\begin{scope}[anchor=west]
  \reprzg{0/1,2/0,2/0} 
  \node at (0.4,-0.4){$\zcfg{\w022}$};
  \draw[thick, -latex] (2,0.25) -- ++(1,0);
  \node at (2.25,0.5) {\instr{s}};
\end{scope}
\begin{scope}[shift={(3.5,0)}, anchor=west]
  \reprzg{1/0,2/0,2/0}
  \node at (0.4,-0.4){$\zcfg{122}$};
\end{scope}
\begin{scope}[shift={(0,-2)}, anchor=west]
  \reprzg{1/0,1/1,2/0}
  \node at (0.4,-0.4){$\zcfg{1\w12}$};
  \draw[thick, -latex] (2,0.25) -- ++(1,0);
  \node at (2.25,0.5) {\instr{p}};
\end{scope}
\begin{scope}[shift={(3.5,-2)}, anchor=west]
  \reprzg{1/0,2/0,2/0}
  \node at (0.4,-0.4){$\zcfg{122}$};
\end{scope}
\begin{scope}[shift={(7,-2)}, anchor=west]
  \reprzg{1/0,1/1,2/0}
  \node at (0.4,-0.4){$\zcfg{1\w12}$};
  \draw[thick, -latex] (2,0.25) -- ++(1,0);
  \node at (2.25,0.5) {\instr{q}};
\end{scope}
\begin{scope}[shift={(10.5,-2)}, anchor=west]
  \reprzg{1/1,0/0,2/1}
  \node at (0.4,-0.4){$\zcfg{\w10\w2}$};
\end{scope}
\begin{scope}[shift={(7,0)}, anchor=west] 
  \reprzg{1/0,2/1,2/0} 
  \node at (0.4,-0.4){$\zcfg{1\w22}$};
  \draw[thick, -latex] (2,0.25) -- ++(1,0);
  \node at (2.5,0.5) {\instr{t}};
\end{scope}
\begin{scope}[shift={(10.5cm,0)}, anchor=west]
  \reprzg{1/1,1/0,2/1}
  \node at (0.4,-0.4){$\zcfg{\w11\w2}$};
\end{scope}
\end{tikzpicture}
\caption{\label{fig:rules in z}
Instructions in $z$-representation described in Table \ref{tab:cases} on 
configuration examples.}
\end{figure}

As briefly mentioned previously, the stable part of a generalized configuration is represented
as a ``ket'' in $z$-representation ; it contains the values of $z(x)$ for all sites. 
Indexed values are repeated as many times as written. The maximal stable configuration
for $L=4$ is thus $\zcfg{2_4}=\zcfg{2222}$. The units in $z$-representations are often called
\emph{particles} because their number is conserved, except at~$x=1$. Waiting particles
in a generalized configuration are represented with the waiting particles listed at the 
left-hand side of the ``ket'' of the stable part, or when convenient, as dots above the number
of particles on the same site : $a_1\zcfg{222}=\zcfg{\w222}$. 

In addition, we introduce a third representation, denoted by $g$, which combines
advantages of the height representation and the slope representation.
For a stable configuration in $\stable$, $g$ is defined as 
\begin{equation}
g(x)=h(x)-\underbrace{(L+1-x)}_{h_{\min(x)}}= \sum_{x'=x}^{L}(z(x')-1).
\label{eq:g}
\end{equation}
The $g$-representation is obtained by removing from the configuration height $h$ all 
the grains below the configuration $c_{\min}$ of uniform slope $1$ 
(the height profile of which is $h_{\min}(x)=L+1-x$).
This change of variable is properly justified by the results of Section \ref{subsect:preliminaries rec conf}.
We refer to the units in the $g$-representation as \emph{stones}. 
To avoid any confusion, when dealing with generalized configurations $c\sim(w,z)\in\general$, 
we always refer to the configuration using the representation $z$ and mostly use the $g$-representation
at the graphical level.
The graphical $g$-representation of any $c\sim(w,z)\in\general$ is obtained simply as follows:
pretend you can settle all the waiting units of $c$ so that you form a new configuration $\bar c\sim(0,z+w)$ 
(which might not be an element of $\stable$)
then use Equation \eqref{eq:g} to build the graphical $g$-representation of $\bar c$.
To finally obtain $c$, 
transform at each site $1\leq x \leq L$ the $w(x)$ upper stones into waiting stones.
We prove in the Section~\ref{sect:classes recur}
that this convention is perfectly adapted to the study of the
stationary state, since all generalized configurations encountered during 
the stabilization process are unambiguously represented.

In the $g$-representation, elementary instructions of Table \ref{tab:cases} are represented 
by simple modifications.
The instruction \instr{p} settles a waiting stone onsite.
The instruction \instr{s} settles a waiting stone directly in contact with a stone 
(waiting or stable) on its right.
Toppling instructions \instr{q} and \instr{t} at a site $x$ 
move a waiting stone to the lowest empty position at the right-hand side site $x+1$
and the upmost stable stone on the nearest left site, $x-1$, is activated, if it exists.
More specifically, instruction \instr{t} is the only one to move a stone to the right \emph{and downward}. 
For $x=L$, the waiting stone is removed from the system. All other situations preserve the number
of stones. 
An example for each situation is given in Figure \ref{fig:stones}.

\begin{figure}[ht!]
\centering
\begin{tikzpicture}
\begin{scope}[anchor=west]
  \reprgg{1/1,2/0,1/0} 
  \node at (0.3,-0.4){$\zcfg{\w022}$};
  \draw[thick, -latex] (2,0.25) -- ++(1,0);
  \node at (2.25,0.5) {\instr{s}};
\end{scope}
\begin{scope}[shift={(3.5,0)}, anchor=west]
  \reprgg{2/0,2/0,1/0}
  \node at (0.3,-0.4){$\zcfg{122}$};
\end{scope}
\begin{scope}[shift={(0,-2)}, anchor=west]
  \reprgg{2/0,1/1,1/0}
  \node at (0.3,-0.4){$\zcfg{1\w12}$};
  \draw[thick, -latex] (2,0.25) -- ++(1,0);
  \node at (2.25,0.5) {\instr{p}};
\end{scope}
\begin{scope}[shift={(3.5,-2)}, anchor=west]
  \reprgg{2/0,2/0,1/0}
  \node at (0.3,-0.4){$\zcfg{122}$};
\end{scope}
\begin{scope}[shift={(7,-2)}, anchor=west]
  \reprgg{2/0,1/1,1/0}
  \node at (0.3,-0.4){$\zcfg{1\w12}$};
  \draw[thick, -latex] (2,0.25) -- ++(1,0);
  \node at (2.25,0.5) {\instr{q}};
\end{scope}
\begin{scope}[shift={(10.5,-2)}, anchor=west]
  \reprgg{1/1,1/0,1/1}
  \node at (0.3,-0.4){$\zcfg{\w10\w2}$};
  \draw (2*0.5-0.25, 1.2) edge [thick, out=0, in=120, ->] (1.25, 1);
\end{scope}
\begin{scope}[shift={(7,0)}, anchor=west] 
  \reprgg{3/0,2/1,1/0} 
  \node at (0.3,-0.4){$\zcfg{1\w22}$};
  \draw[thick, -latex] (2,0.25) -- ++(1,0);
  \node at (2.25,0.5) {\instr{t}};
\end{scope}
\begin{scope}[shift={(10.5,0)}, anchor=west]
  \reprgg{2/1,2/0,1/1}
  \node at (0.3,-0.4){$\zcfg{\w11\w2}$};
  \draw (2*0.5-0.25, 1*0.5+2*0.5) edge [thick, out=45, in=90, ->] (3*0.5-0.25, 0*0.5+2*0.5);
\end{scope}
\end{tikzpicture}
\caption{\label{fig:stones}
Instructions in $g$-representation applied on the same initial configurations
as in Figure~\ref{fig:rules in z}. 
Note that waiting stones at $x'$ coincide with waiting particles at $x'$.
The movement of a stone caused by a toppling is represented as an arrow.}
\end{figure}

The height levels of the $g$-representation where there are stones are called \emph{rampart}.
They organize into \emph{merlons} and \emph{crenels} alternatively.
Merlons and crenels are made of consecutive stones and empty positions
respectively (see Figure \ref{fig: crenels merlons}).
A crenel always has a merlon to its right and a merlon, or the wall at $x=0$, to its left.
Topplings always occur at the rightmost position of a merlon. 
If the stone stays in the same rampart and moves from $x$ to $x+1$, three outcomes are possible~:
(i) it creates a new merlon at $x+1$ ($x+2$ is empty), 
(ii) or it displaces the position of an elementary merlon (no stones at $x-1$ and $x+2$),
(iii) or merges to the merlon standing on the right (stone at $x+2$).
If it goes down one level, it
extends the rampart on which it was standing by one stone to the right.

\begin{figure}[ht!]
\centering
\begin{tikzpicture}
\begin{scope}
  \reprgg[black]{2/1,2/0,2/0,1/1,1/0,1/1,1/0,2/0,1/1,1/0,0/1}
\end{scope}
\draw[dashed](0,0.5) -- ++(10.5,0);
\draw[dashed] (0,1) -- ++ (10.5,0);
\node[font = \itshape] at (6,1.5) {rampart};
\begin{scope}[xshift = 6 cm, yshift = 0.5 cm]
  \reprg{1,1,1,1,0,1,0,1,1,0,0}{0.5}{black} 
  \node at (2.7,1.5){crenels};
  \draw[->] (2.75,1.2) -- (2.25,0.4);
  \draw[->] (2.75,1.2) -- (3.25,0.4);
  \node at (3,-1.2) {merlons};
  \draw[->] (2.5,-0.95) -- (1.25,-0.1);
  \draw[->] (2.5,-0.95) -- (2.75,-0.1);
  \draw[->] (2.5,-0.95) -- (3.5,-0.1);
\end{scope}
\end{tikzpicture}
\caption{\label{fig: crenels merlons}  Left: $g$-representation of 
the configuration $\zcfg{\w111\w10\w101\w11\w1}$.
Waiting stones are represented as gray squares and stable stones as white squares.
Right: we isolate the rampart at level $y=2$,
which is composed of $3$ merlons and $2$ crenels.} 
\end{figure}

\section{Description of the stabilization}\label{sect:def stab}
In this Section, we study the steady state of the Oslo model. 
We take advantage of the important result that the steady state is the
statistical state obtained after injection of a unit at $x=1$ to the maximal configuration
$\zcfg{2_L}$ and stabilization, as proved by Dhar in the Ref.~\cite{dhar_steady_2004}. In the same article,
Dhar also showed that the steady state is reached after at most $L(L+1)$
injections starting from an initial stable configuration,
the scaling of which, $L^2$, is reminiscent of the relaxation time in diffusive systems.

In this Section, we introduce two complementary ways to describe the stabilization process:
Decision trees and Stacks of Instructions.
These latter objects have proven essential in the study of other stochastic sandpile models such as ARW
\cite{hoffman_density_2024}.

\subsection{Evolution operators and decision trees}
\label{sec:paths}

The stochastic nature along the stabilization process generates distributions of configurations which are element of the vector space $\stat:=\textrm{span}(\general)$.
Each element of $\stat$ is called a generalized state and is a sum of positive weights decomposed over the elements of $\general$.
Each weight is interpreted as a probability of being in its associated configuration.
A stable state has only non zero weights on stable configurations.

\paragraph{Example.}
Let us consider the configuration~$\zcfg{111}$ and add a waiting unit at $x=2$, 
the first steps of
the stabilization are the following generalized states
\[ \zcfg{1\w11} \to p\zcfg{121} + q\zcfg{\w10\w1} \to p\zcfg{121} +
 pq\zcfg{\w102} + q^2\zcfg{\w1\w0\w0} \to\cdots \]
From this example, we observe that the order in which 
the instructions are executed modifies the sequence of statistical states.
This is, however, true along the stabilization process but the final state, which is
a probability distribution on stable configurations, does not depend on this order \cite{dhar_steady_2004}.

We now give a mathematical definition of the arrows from the above example. Let us define
\begin{definition}[Evolution operators] \label{def:evol}
An evolution operator~$\evol$ is a Markov operator acting on statistical states, 
defined by its action on $c\in\general$ as follows
\begin{itemize}
    \item $\evol$ selects a waiting unit $a_x$ in $c$;
    \item $\evol$ executes the corresponding instructions depending on $z(x)$ 
          given in Table \ref{tab:cases}. \\
          The result is the sum of possible generalized configurations weighted by their probabilities of occurrence.
\end{itemize} 
On the set of stable configurations, $\evol$ is the identity. In other words, if there
are no waiting units, the evolution operator does not modify the configuration~$c$. 
The set of evolution operators is denoted by~$\evolution$.
\end{definition}

The role of an evolution operator~$\evol$ is to determine the order in which 
waiting units, or sites, equivalently, are activated and to execute locally all the possible instructions at once.
There exists an operator $\evol$ for every possible sequence of choices along an avalanche. 
The repeated action of an evolution operator builds a decision, or evolution, tree weighted with probabilities.

Continuing the example of the previous section, we write 
$\evol\zcfg{1\w11}=p\zcfg{121}+q\zcfg{\w10\w1}$,
to show that the evolution operator $\evol$ is a mathematical formalization of the arrows we have used so
far to express a sequence of states.
It must be noted that a given configuration can appear at several nodes in the evolution tree.

Evolution trees can in principle start from any generalized configuration. 
However, in our study, we only study evolution trees starting with generalized configurations
of the form $a_1c$, where $c\in\stable$. In most cases, the initial generalized configuration is
$a_1\zcfg{2_L}$, as its stabilisation was shown in \cite{dhar_steady_2004} to give the stationary state of the model.
An example of an evolution tree is displayed in the Figure~\ref{fig:tree L=2}.

\begin{definition}[Path] \label{def:path}
    An avalanche, \textit{i.e.} a sequence of \emph{generalized configurations}, 
    is a \emph{path} in an evolution tree. 
    A sequence of configurations that can be obtained as a path in an evolution tree is \emph{legal},
    any other sequence of configurations is \emph{non-legal}.
\end{definition}
The notion of \emph{legal} or \emph{non-legal} is an intuitive adjective that appears systematically in all the representations of the Oslo model evolution.
In simple terms, any description of the system that follows the model rules summarized in Table \ref{tab:cases} is legal, all the rest is non-legal.

\paragraph{Continuing example}
In the previous \textbf{Example}, a possible configuration path
starts with the sequence
\[ \zcfg{1\w11},\quad \zcfg{\w10\w1},\quad \zcfg{\w102}, \quad \zcfg{202}. \]
It is equivalent to the starting generalized configuration~$\zcfg{1\w11}$ followed
by the sequence of instructions
\instrsub q2, \instrsub p3, \instrsub p1, where \instrsub ux denotes an instruction
of kind \instr{u} at the site~$x$. 

As previously mentioned, the model is set in a driven-dissipative setup where waiting particles are added one at a time, and after stabilization of the previous injection, on the first site.
We are therefore interested in stablizing all configurations of the form $a_1c$ with $c\in\stable$. 
The leaves of an evolution tree generated by $\evol\in\evolution$ with root $a_1c$, 
are the stable configurations~$c'$ that result
from avalanches starting with~$a_1c$. 
The total probability of all paths from 
$a_1c$ to $c'$ is written as $\avalanche(c,\,c')$ and does not depend on the choice of $\evol$ as the model is Abelian \cite{dhar_steady_2004}. This defines the linear operator~$\avalanche$
as the \emph{avalanche matrix} for the Oslo model. 
For any $\evol\in\evolution$, we therefore have the equality

\begin{equation*}
    \evol^\infty a_1=: \avalanche
\end{equation*}
The symbol $\evol^\infty$ denotes the repeated application of the evolution operator until stabilization is
complete. This is well defined, since all unstable configurations stabilize after a finite number of instructions, as shown in Ref.~\cite{dhar_steady_2004}.
Moreover, evolution operators are irreversible since topplings move grains to the right
and no instructions move grains to the left in the model.
On the contrary, $\avalanche$ is reversible, as it contains also the injection part.
An example of evolution tree is provided in the Figure~\ref{fig:tree L=2} in the case $L=2$. 

Another remarkable result of \cite{dhar_steady_2004}, which is at the heart of the present paper, states that if 
the starting configuration is $a_1\zcfg{2_L}$, the stabilized state is the stationary state~$\psi$  that satisfies
\begin{equation}\label{eq:stst relation}
    \evol^\infty(a_1\psi)=\psi
\end{equation}
The associated evolution tree is said to be \emph{maximal}.
As the avalanche matrix~$\avalanche$ is a Markov matrix, the stationary state
is the statistical state with eigenvalue~1.

\begin{definition}[Recurrent configurations]
Configurations with a non-zero probability in~$\stst$ are called \emph{recurrent configurations}
while the others are called
\emph{transient}. 
The set of recurrent configurations is 
denoted by~$\recur$ and is a strict subset of~$\stable$.
\end{definition}

\begin{figure}[ht!]
    \centering
     \scalebox{0.9}{
        \begin{forest}
        for tree={
    every leaf node={rounded corners=2pt, text=blue, draw=black},
    l sep=6mm,
    edge={-Straight Barb},
    EL/.style = {edge label={node[midway, fill=white, inner sep=1pt, anchor=center]{\instr{#1}}},},
    }
    [$\A1\zcfg{22}$
      [$\A2\zcfg{12}$, EL = t
        [$\A1 a_2\zcfg{11}$, EL = t
            [$\A2\zcfg{21}$,EL=p
              [$\zcfg{22}$,EL=p]
              [$a_1\A2\zcfg{20}$,EL=q
                [ $\A1\zcfg{21}$,EL=s
                  [$\A2\zcfg{11}$,EL=t
                    [$\zcfg{12}$,EL=p]
                    [$a_1\A2\zcfg{10}$,EL=q
                      [$\A1\zcfg{11}$, EL = s
                        [$\zcfg{21}$,EL=p]
                        [$\A2\zcfg{01}$,EL=q
                          [$\zcfg{02}$,EL=p]
                          [ $a_1\A2\zcfg{00}$,EL=q
                            [$\A1\zcfg{01}$, EL = s
                              [$\zcfg{11}$, EL = s]
                            ]
                          ]
                        ]
                      ]
                    ]
                  ]
                ]
              ]
            ]
            [$a_2\A2\zcfg{01}$,EL=q
              [ $\A2\zcfg{02}$ ,EL=p
                [$\A1a_2\zcfg{01}$, EL = t
                  [$\A2\zcfg{11}$, EL = s
                    [$\zcfg{12}$,EL=p]
                    [$a_1\A2\zcfg{10}$,EL=q
                      [$\A1\zcfg{11}$, EL = s
                        [$\zcfg{21}$,EL=p]
                        [$\A2\zcfg{01}$,EL=q
                          [$\zcfg{02}$,EL=p]
                          [ $a_1\A1\zcfg{00}$,EL=q
                            [$\A1\zcfg{01}$, EL = s
                              [$\zcfg{11}$, EL = s]
                            ]
                          ]
                        ]
                      ]
                    ]
                  ]
                ]
              ]
              [$a_1a_2\A2\zcfg{00}$,EL=q
                [$\A1a_2\zcfg{01}$, EL = s
                  [$\A2\zcfg{11}$, EL = s
                    [$\zcfg{12}$,EL=p]
                    [$a_1\A2\zcfg{10}$,EL=q
                      [$\A1\zcfg{11}$, EL = s
                        [$\zcfg{21}$,EL=p]
                        [$\A2\zcfg{01}$,EL=q
                          [$\zcfg{02}$,EL=p]
                          [ $a_1\A2\zcfg{00}$,EL=q
                            [$\A1\zcfg{01}$, EL = s
                              [$\zcfg{11}$, EL = s]
                            ]
                          ]
                        ]
                      ]
                    ]
                  ]
                ]
              ]
            ]
          ]
        ]
      ] 
    \end{forest}
  }   
  \caption{\label{fig:tree L=2}
  A maximal evolution tree for $L=2$ and a given $\evol\in\evolution$. 
  The waiting particle executing an instruction is highlighted in red.
  We observe that there are $N_2=5$ different recurrent configurations with 
  $\recur(2)=\{\zcfg{22},\zcfg{12},\zcfg{21},\zcfg{02},\zcfg{11}\}$.
  Notice that all trees with root $a_2\zcfg{11}$ are the same.}
\end{figure}

Despite its apparent simplicity, the dynamics of the Oslo model requires a large number of operations. 
It was shown that the number of topplings in one
avalanche scales up to $L^3$ \cite{dhar_steady_2004}, 
such that the number of operations needed to compute a single row 
of the matrix~$\avalanche$ is quite large
and is bounded from above by $2^{\mathrm{cst}.L^3}$.
The properties of the stationary state therefore have a computational cost directly
inherited from this observation. 
We can find exact numerical approach of this problem in \cite{corral_calculation_2004} and \cite{pradhan_probability_2006,pradhan_sampling_2007}
where system up to $L=8$ and $L=12$ were respectively investigated. 

\subsection{Stacks of Instructions}
The Abelian nature of the stabilization is calling
for an invariant representation of the dynamics with respect to the choice of $\evol\in\evolution$.
This is precisely what the Stacks of Instructions are designed for. 
This kind of representation has been used in several published works under different names, depending on the model
and the community. 
For instance, in the literature of the ARW it goes as the \emph{sitewise representation}
(see e.g. \cite{hoffman_density_2024}).

\begin{definition}[Stacks of Instructions]\label{def:stacks}
    We call \emph{stack of instructions at~$x$} the ordered sequence of instructions performed at a given site~$x$ 
    during stabilization. All stacks of instructions form together the \emph{Stacks of Instructions} (\SI{}).
    Geometrically, we represent the \SI{} on a square lattice, by coloring 
    the squares $(x,\,y)$ respecting $1\leq x\leq L$ and $1\leq y\leq n(x)$ where $n(x)$ is the length of the 
    stack at~$x$, and the colors are elements from $\{\instr p,\,\instr q,\,\instr s,\,\instr t\}$.
    An example of \SI{} is provided in the Figure~\ref{fig:stacks}.
    
    Similarly, we define \emph{Stacks of Random Instructions} (\SRI{}) and \emph{Stacks of Toppling Instructions} 
    with the same rules, but these Stacks only account for the random instructions (\instr p and \instr q) and toppling instructions (\instr q and \instr t) respectively. 
    A \SI{} can therefore
    be transformed into a \SRI{} (\STI{}) by considering, for each stack, the random 
    (toppling) instructions and keeping the same relative order between them.
    Examples of \SRI{} and \STI{} can be found in the Figure~\ref{fig:stacks}.
\end{definition}

\noindent
\begin{figure}[ht!]
    \centering
    \begin{forest}
       for tree={
         l sep=2.5mm,
         edge={-Straight Barb}, 
         EL/.style = {edge label={node[midway, inner sep=2.5pt, anchor=south] {#1}}},
         grow=east}
      [$\A1\zcfg{22}$
      [${\color{teal}a_2}\zcfg{12}$, EL = {\instrsub[purple]t1}
      [$\A1a_2\zcfg{11}$, EL = {\instrsub[teal]t2}
      [${\color{teal}a_2}\zcfg{21}$, EL= {\instrsub[purple]p1}
      [$a_1{\color{teal}a_{2}}\zcfg{20}$,EL = {\instrsub[teal]q2}
      [$\A1\zcfg{21}$,  EL = {\instrsub[teal]s2}
      [${\color{teal}a_2}\zcfg{11}$, EL = {\instrsub[purple]t1}
      [$\zcfg{12}$, EL = {\instrsub[teal]p2}]]]]]]]]
    \end{forest}
    \caption{\label{fig:path tree L=2}
    The second path from the left between the root and a leaf in the tree of Figure \ref{fig:tree L=2}.}
    
\end{figure}

\begin{figure}[ht!]
\centering
\begin{tikzpicture}
  \begin{scope}
    \xaxis{1.5}{1,2}
    \reprh{3,4}{0.5}{black}
    \node at (0.25,0.25) {\instr[purple]t};
    \node at (0.25,0.75) {\instr[purple]p};
    \node at (0.25,1.25) {\instr[purple]t};
    \node at (0.75,0.25) {\instr[teal]t};
    \node at (0.75,0.75) {\instr[teal]q};
    \node at (0.75,1.25) {\instr[teal]s};
    \node at (0.75,1.75) {\instr[teal]p};
    \node[rectangle,draw] at (1.25,-0.75) {\SI{}};
  \end{scope}

  \begin{scope}[xshift = 4cm]
    \xaxis{1.5}{1,2}
    \reprh{1,2}{0.5}{black}
    \node at (0.25,0.25) {\instr[purple]p};
    \node at (0.75,0.25) {\instr[teal]q};
    \node at (0.75,0.75) {\instr[teal]p};
  \node[rectangle,draw] at (1.25,-0.75) {\SRI{}};
  \end{scope}
  
  \begin{scope}[xshift = 8cm]
    \xaxis{1.5}{1,2}
    \reprh{2,2}{0.5}{black}
    \node at (0.25,0.25) {\instr[purple]t};
    \node at (0.25,0.75) {\instr[purple]t};
    \node at (0.75,0.25) {\instr[teal]t};
    \node at (0.75,0.75) {\instr[teal]q};
  \node[rectangle,draw] at (1.25,-0.75) {\STI{}};
  \end{scope}
\end{tikzpicture}
\caption{\label{fig:stacks}
  (Left) Stacks of Instructions (\SI{}) of the path from Figure \ref{fig:path tree L=2}. 
  (Middle) Stacks of Random Instructions (\SRI{}), where instructions \instr{s} and \instr{t} have been removed.
  (Right) Stacks of Toppling Instructions (\STI{}), the \SI{} where only \instr{q} and \instr{t} are kept.}
\end{figure}

\begin{definition}[Paths and legal Stacks]\label{def:path legal stack}
    Given an evolution tree generated by $\evol\in\evolution$, we can generate, for each path from the root to a tree leaf, 
    the corresponding \SI{}.
    It suffices to stack the instructions at each site along the path, adding on top of the corresponding stack the latest
    instruction executed.
    Given an initial configuration $c\in\general$, we say that a \SI{} (\SRI{}/\STI{}) $S$ is \emph{legal} 
    if and only if there exists an operator 
    $\evol\in\evolution$ that generates $S$ starting from $c$.
    Otherwise the \SI{} (\SRI{}/\STI{}) is \emph{non-legal}.
\end{definition}

\begin{prop}
\label{prop:unique stacks}
Inside the evolution tree generated by an evolution operator $\evol\in\evolution$,
two different paths cannot generate the same \SRI{}. Consequently two different paths cannot generate the same \SI{}.
\end{prop}

\begin{proof}
In an evolution tree, branchings only occur when a site~$x$ with $z(x)=1$ is activated
and the branches correspond to instructions \instr{p} or \instr{q}. 
The only way two paths can differ is by trading at least a \instr{q} for a \instr{p},
or the opposite. Therefore, they cannot produce the same \SRI{}, and a fortiori the same \SI{} by inclusion. 
\end{proof}

Let us remark that, up to now, the instructions and the order were generated altogether by the choice of $\evol\in\evolution$.
We decouple these two contributions, using the \SRI{} to encode the random instructions performed at each site during stabilization, and introduce \emph{odometers} to specify the order in which they are performed.

\begin{definition}[Odometer]\label{def:odometer}
    An \emph{odometer} $\mu:\intrange{1}{L}\to\N$ is a function specifying
    the number of instructions performed at each site in $\intrange{1}{L}$ by waiting units.
    They can be defined on Stacks of Instructions of any kind (\SI{}, \SRI{} and \STI{}).
\end{definition}

As argued in Proposition \ref{prop:unique stacks}, \SRI{} are sufficient to distinguish paths, and so we mostly consider odometers defined on \SRI{}.

\begin{definition}[Path, \SRI{} and legal odometers]\label{def:path, SRI and legal odometers}
    A path in an evolution tree with root $c\in\general$ is equivalently seen 
    as a sequence of configurations $(c_t)_t$, or as a legal \SRI{}, denoted $S$,
    plus a sequence of increasing odometers $(\mu_t)_t$.
    The sequence of odometers $(\mu_t)_t$ is said to be \emph{legal} if and only if each transition $c_t\to c_{t+1}$
    corresponds to executing one random instruction, let's say at site $x$, as defined in Table \ref{tab:cases},
    and $\mu_{t+1}(x') = \mu_t(x') +\mathbf1_{\{x'\}}(x)$ where $\mathbf1_U(x)$ is the indicator function on the set $U$, with the initial condition $\mu_0(x')=0$.
    Otherwise the sequence is \emph{non-legal}.
    This construction supposes that all legal deterministic instructions \instr s and \instr t are implicit.
    They can be inferred from the model rules and are executed instantly whenever they must be involved.
    We call the last element of the odometer sequence associated to a stabilizing path,
    meaning one leading to a stable configuration $r\in\stable$, the stabilizing odometer of $r$, and denote it by $\mu_r(x)$.
    
    By construction of the evolution operator (Definition \ref{def:evol}),
    there exists at least one $\evol\in\evolution$ which generates both $S$ and $(\mu_t)_t$.
    Supposing we only know the initial configuration $c\in\general$ and $S$ to be a legal \SRI{},
    we say abusively that an operator $\evol\in\evolution$ generates on $S$ a legal sequence of odometers $(\mu_t)_t$ 
    and a legal sequence of configurations $(c_t)_t$.
\end{definition}

\begin{prop}\label{prop:SI defines path}
Consider the evolution tree generated by $\evol\in\evolution$ with root $c\in\general$ and a path from $c$ to a leaf $r\in\stable$.
This path defines a legal \SI{} that we denote $S$.
Consider also the evolution tree obtained from a different evolution operator $\evol'\in\evolution$ with the same root $c$. Then, this new tree contains a path from $c$ to $r$ that defines the same legal \SI{} $S$.
\end{prop}

This result is a special case of deterministic Abelian sandpile models, 
such as the BTW \cite{bak_self-organized_1987}. 
The determinism here comes from the specification, in advance, of the instructions through the \SI{}.
We follow the same approach as Dhar (see e.g. the review \cite{dhar_theoretical_2006}).

\begin{proof}
We consider $c$, $r$, $\evol$ and $S$ as in the Proposition.
By construction, $\evol$ generates a legal sequence of configurations $(c_t)_t$.
Each step $c_t\to c_{t+1}$ corresponds to a single instruction of $S$.
If there is only one step, there is nothing to prove, so we consider the number of steps to be at least 2.
At the path level, consider $\evol'\in\evolution$ such that it differs only by one 
\emph{elementary transposition} of the order of activation defined by $\evol$ on $S$ and denote by $(c'_t)_t$ the associated sequence of configurations.
Then, there is a time $t_1$ such that 
$\evol$ activates site $x$ of $c_{t_1}$ and then site $y$ of $c_{t_1+1}$, 
whereas $\evol'$ activates respectively site $y$ in $c'_{t_1}$ and then site $x$ in $c'_{t_1+1}$.
The only possibility for this situation to happen is if $c_{t_1}=c'_{t_1}$ have at least one waiting 
unit at sites $x$ and $y$.
After executing the two instructions, we can check that both $\evol$ and $\evol'$ lead to the same 
configuration $c_{t_1+2}=c'_{t_1+2}$. 
Indeed, if $x=y$ or if $x\neq y$ and the two successive instructions settle the particles, this is obvious.
Otherwise $x\neq y$ and the instructions involve at least one toppling.
Remark that both updates are independent as $x\neq y$ and the instructions depend on the stable part which 
can only be modified after the execution of an instruction.
So the instructions at $x$ can not modify the stable part at $y$ and vice-versa.
Moreover, a toppling sends a constant number of particles out of the site and depletes the stable part 
and the waiting units of the site by a constant number (minus one for each fields). 
We must therefore have the equality $c_{t_1+2}=c'_{t_1+2}$.
Finally, as the order specified by $\evol$ and $\evol'$ coincide anywhere else, they both lead to the same
stable configuration.

We now prove that we can "synchronise", step by step, the operator $\evol$ to $\evol'$.
To do so, we build a sequence of operators $(\evol_k)_{1\leq k\leq n}$
in a controlled way, where $\evol_1:=\evol$ 
and $\evol_n=\evol'$.
Each step of the procedure corresponds to constructing a 
new operator $E_{k+1}$ based on $E_{k}$ by performing
only elementary transpositions.
Since these operations ensure that both $E_{k}$ and $E_{k+1}$ lead to the same stable configuration 
given $S$, both the final configuration and the~$\SI{}$ are invariant during this process.

Let us denote by $(c_t^k)_t$ 
the sequence of configurations defined by $\evol_k$ on $S$,
and such that $c_1^k:=c$ for all $k$.
At each step, we build $\evol_{k+1}$ from $\evol_{k}$ finding the first element at time $t_1$
such that $c_{t_1}^k\neq c'_{t_1}$. 
Let us denote by $x$ the site at which the instruction has been executed between $c'_{t_1-1}$ and $c'_{t_1}$.
From the hypothesis, $c_{t_1-1}^k = c'_{t_1-1}$, and so we know that a waiting unit lives on site $x$ also in $c_{t_1-1}^k$.
This waiting unit at $x$ will necessarily be activated by $\evol_k$ at a time $t_2>t_1$. 
For all times $t_1\leq t\leq t_2$, all corresponding $c_t^k$ have a waiting unit at $x$.
It suffices now to perform the sequence of elementary transposition $t_2\to t_2-1 \to... \to t_2-(t_2-t_1)=t_1$,
meaning we replace the $t_1$-th instruction executed by $E^k$ by the $t_2$-th one.
This replacement defines the sequence $(c_t^{k+1})_t$ such that $c_{t}^{k+1}=c'_{t}$ for all $t\leq t_1$
and defines $\evol_{k+1}$.
It is clear that this procedure can be continued until we get to $\evol'$ and so, by induction on $k$,
we get the desired result.
\end{proof}

Proposition \ref{prop:SI defines path} says that the model is Abelian at the deterministic level.
This result implies that stabilization with the stochastic rules is also Abelian: it suffices
to sum up all possible realisations of the \SI{}.
Given an initial configuration $c\in\general$, if we know in advance the instructions to perform at each sites,
and that there is at least one legal way to execute all the instructions legally, then any legal order of activation,
starting from $c$, leads to the exact same final configuration.
In fact, the same result also applies to the \SRI{}, using Proposition \ref{prop:unique stacks}
and the convention that all deterministic instructions are executed implicitly.

\subsection{The toppling function}
The \SI{} contains all the information on the path instructions, and we decompose the latter into different 
contributions: a (strictly) stabilizing and a current part. The latter can be encoded,
regardless of its stochastic or deterministic nature, into the \emph{toppling function}. 

\begin{prop}[Toppling function]\label{prop:toppling}
Let us consider a path in an evolution tree and two generalized configuration~$c_1$ and $c_2$
such that the configuration $c_2$ occurs after $c_1$ along the path. 
We denote by $z_i$ and $w_i$ ($i=1,2$) the slope of the stable part and the number of waiting 
units of $c_i$ respectively. 

Let us define the \emph{toppling function}~$\toppling[12]$ such that
$\toppling[12](0)=0$, 
$\toppling[12](L+1)=\toppling[12](L)$
and for $1\leq x\leq L$, $T_{12}(x)$ is 
the number of toppling instructions (\instr q or \instr t) performed
from $c_1$ to $c_2$ at site~$x$. 
Then the $z$-representations of $c_1$ and~$c_2$ are related by
\begin{equation}
  \forall x\in\intrange{1}{L},\; z_2(x) + w_2(x) = z_1(x) + w_1(x) + \Delta\toppling[12](x)
  \label{eq:toppling z}
\end{equation}
where $\Delta f(x) = f(x+1) + f(x-1)-2f(x)$
is the discrete Laplacian. 
\end{prop}

\begin{proof}
Suppose that~$c_2$ immediately follows~$c_1$ along the path, after a toppling instruction 
(\instr{q} or \instr{t}) at site~$x$. The
corresponding toppling function is $\toppling[12]$ such that $\toppling[12](x)=1$ and
$\toppling[12](x')=0$ for $x'\neq x$. We also know from Table~\ref{tab:cases} that 
\begin{align*} 
z_2(x-1)+w_2(x-1)&=z_1(x-1)+w_1(x-1)+1,\\
z_2(x)+w_2(x)&=z_1(x)+w_1(x)-2, \\
z_2(x+1)+w_2(x+1)&=z_1(x+1)+w_1(x+1)+1.
\end{align*}
which proves the announced result except for $x=1$ and $x=L$. 
The other instructions (\instr{s} and \instr{p}) leave the sum $z+w$ invariant and therefore correspond to
the Laplacian of the null function. Since the formula is linear, it remains true after several instructions. 

To extend its validity to sites $x=1$ and $x=L$, we only have to
define the values of~$\toppling[12]$ at sites $x=0$ and $x=L+1$ in such a way that 
the equality~\eqref{eq:toppling z} remains true. For a single toppling at $x=1$, we get 
\[ \toppling[12](0)=2\toppling[12](1)-\toppling[12](2)+z_2(1)+w_2(1)-z_1(1)-w_1(1)=0\]
and for a single toppling at $x=L$ 
\[ \toppling[12](L+1)-\toppling[12](L)=\toppling[12](L)-\toppling[12](L-1)+z_2(L)+w_2(L)-z_1(L)-w_1(L)=0\]
Therefore, by defining~$\toppling[12](0)=0$ and $\toppling[12](L+1)=\toppling[12](L)$, formula
\eqref{eq:toppling z} is always verified.
\end{proof}

Since the difference between $c_1$ and $c_2$ is independent of the path, 
the toppling function~$\toppling[12]$ is also independent of the path. 
In other words, Proposition \ref{prop:toppling} proves the invariant nature of the toppling function. 
Any path from $a_1\zcfg{2_L}$ to some fixed $r\in\recur$ must have exactly the same toppling functions.
We can then denote by $\toppling[c_1,\,c_2]$ the toppling function from~$c_1$ and $c_2$ without ambiguity
and remark that the toppling function describes the profile of the \STI{} associated 
to any path from $c_1$ to $c_2$.  
The proof of Proposition~\ref{prop:toppling} uses \emph{transitivity} property of the toppling function 
that states
\[ \toppling[c_1,\,c_2]+\toppling[c_2,\,c_3] = \toppling[c_1,\,c_3] \]
for three configurations~$c_1$, $c_2$ and~$c_3$ of the same path occurring in this order.

Whenever the configuration~$c_1$ is the particular $a_1\zcfg{2_L}$ and $c_2=r\in\recur$, we simply denote 
$\toppling[r]$ the associated toppling function. 
The path shown as an example in the Figure \ref{fig:stacks} has a toppling function such 
that $\toppling(1)=\toppling(2)=2$.

\begin{prop}[Structure of the \SI{} and \STI{}]
\label{prop:struct toppling}
We consider the \SI{} and \STI{} of any path starting from the configuration~$a_1\zcfg{2_L}$
down to a recurrent configuration~$r$. 
They satisfy the following properties
\begin{itemize}
\item[a)] The \STI{} profile, i.e. the toppling function, is concave.
\item[b)] The first row of the \SI{} and \STI{} is made of $L$ instructions~\instr{t}.
\item[c)] In the \SI{}, the stack on site $x\in\llbracket 1,L\rrbracket$ is made,
from the second row (included), of successive pairs \instr{p} and \instr{t} or \instr{q} and \instr{s} 
instructions.
There is potentially an exception for the last instruction of the stack.
\item[d)] For the \SI{}, the last instruction of the stack on site~$x\in\llbracket 1,L\rrbracket$ is 
\[ \begin{cases} \instr{p} & \text{if $z_r(x)=2$} \\
                 \instr{q} & \text{if $z_r(x)=0$} \\
                 \instr{s} \text{ or } \instr{t} & \text{if $z_r(x)=1$}
\end{cases}
\]
\end{itemize}
\end{prop}

\begin{proof}
\begin{itemize}
\item[a)] 
$\toppling[r]$ is concave because the initial configuration is $c_0=a_1\zcfg{2_L}$, so we necessarily have
$z_r(x) \leq z_0(x)+w_0(x)$. From the equation~\eqref{eq:toppling z}, this implies that  $\Delta\toppling\leq 0$. 
\item[b)]
Since the starting configuration has $z(x)=2$, the only possible first instruction, at all $x$, is~\instr{t}. 
Therefore, the injected particle $a_1$ in $a_1\zcfg{2_L}$ is necessarily dissipated from $x=L$
during the stabilization process.
\item[c)] After a stochastic process at $x$, the local slope is either $0$ or $2$. 
After the addition of a waiting unit at $x$, the next instruction is deterministic as specified 
in the Table~\ref{tab:cases}. 
Consequently, the \SI{} rows alternate between rows with \instr{p} and \instr{q} only and rows
with \instr{s} and \instr{t} only.
Each pair contributes for one unit to $\toppling(x)$. 
\item[d)] 
This result is a simple translation of Table~\ref{tab:cases} concerning the last instruction on a site.
\end{itemize}
\end{proof} 

Let us now define two invariants associated to the recurrent configurations.
\begin{definition}[Final domain]
The set of \emph{final} \instr{p} instructions of a \SRI{} is called \emph{final domain}. 
For a given path between $a_1\zcfg{2_L}$ and $r\in\recur$, we define it as 
\begin{equation}
 \final[r] = \big\{ (x, \toppling(x)), \; z(x)=2\big\}.
\label{def:Pr} 
\end{equation}
\end{definition}

\begin{prop}[Invariance of the final domain]
\label{prop:inv final}
The final domain $\final[r]$ of $r\in\recur$ is invariant over all legal \SRI{} leading to $r$: $\final[r]$ is necessarily colored with \instr{p} instructions.
Therefore, the final domain does not depend on the path leading to~$r$.
\end{prop}

\begin{proof}
Straightforward from the Proposition ~\ref{prop:struct toppling}.d :
the $x$ coordinates of the positions in $\final[r]$ 
defines the sites where $z(x)=2$ for $r$, 
and the unique way to do it is by executing a $\instr{p}$ instruction.
\end{proof}

\begin{definition}[Toppling domain]
\label{def:toppling domain}
Consider $r\in\recur$.
We call $\domain[r]$ the \emph{toppling domain}
defined as
\begin{equation}\label{eq:toppling domain}
    \domain[r] = \{(x,t)|~1\leq t\leq \toppling[r](x)-1\}
\end{equation}
where the first coordinate of each position is the site and the second 
the order of the instruction associated to a toppling.

Then every path from $a_1\zcfg{2_L}$ to~$r$ is associated to a \SRI{} which is defined on a constant domain denoted 
by $\domain[r]\cup\final[r]$.
Furthermore, we use the notation $|\domain[r]|$ for the cardinal of $\domain[r]$,
and denote by $|\domain[r]|(x)$ the cardinal of $\domain[r]$ restricted to the site $x\in\intrange{1}{L}$. 
We then have $|\domain[r]|=\sum_{x=1}^L |\domain[r]|(x)$.
\end{definition}

The results of this section invite to consider the \SRI{} as a central object of study for many reasons
(see Figure \ref{fig:sketch RSoI}). 
The \SRI{} naturally splits into a stabilizing part $\final[r]$ and a toppling part $\domain[r]$.
The product of all probabilities $p$ and $q$ associated to instructions 
\instr{p} and \instr{q} in the \SRI{} gives the probability of the path.
We have also a complete description of the topplings as any \instr{p} at a position $v\in\domain[r]$ 
of the \SRI{} is associated to a \instr{t} (cf Proposition \ref{prop:struct toppling}). 
Since the first instruction at each site is \instr{t}, the first row in the \SI{} can be ignored
(cf Proposition \ref{prop:struct toppling}). 
Finally, given a path $a_1\zcfg{2_L}\to r\in\recur$, we are ensured that the stabilizing odometer 
of $r$ is just given by $\mu_r(x):=|\domain[r]\cup\final[r]|(x)$.

\begin{figure}[ht!]
\centering
\begin{tikzpicture}
\draw[thick, black!50, domain=1:2] plot (\x,{0.3+0.15*\x*(8-\x)}) 
    -- ++(0,-2) -- ++(-1,0) -- cycle;
\draw[thick, black!50, domain=2.5:3.5] plot (\x,{0.3+0.15*\x*(8-\x)})
    -- ++(0,-2) -- ++(-1,0) -- cycle;
\draw[thick, fill=black!20, domain=0:4] plot (\x,{0.2+0.15*\x*(8-\x)}) 
    -- (4,0) -- (0,0) ;
\draw[<->]  (0,3) -- (0,0) -- (5,0) node[right] {site};
\node (P) at (2.25,3) {$\final[r]$};
\draw[->] (P) edge[out=225, in=120] (1.5, 1.8);,
\draw[->] (P) edge[out=75, in=110, out=0] (3, 2.55);
\node[rotate=90] at (-0.5,1.5) {random instructions};
\node at (5.25,1.85) {\instr{p} and \instr{q}} edge[out=-90, in=0, ->] (3.5, 1);
\end{tikzpicture}
\caption{\label{fig:sketch RSoI} Sketch of the Stacks of Random Instructions 
of a path from~$a_1\zcfg{2_L}$ to a configuration~$r\in\recur$.
The grey area is exactly the toppling domain $\domain$ which contains the random instructions associated to topplings. 
Its height (thick black line) is given by $\toppling(x)-1$ as the deterministic
initial toppling \instr{t} is not included on each site.
Final \instr{p} instructions, grouped into~$\final[r]$,
are outside of the grey area, because they
do not contribute to the toppling function. 
These \instr{p} are present at all sites $x$ where~$z(x)=2$ in $r$.}
\end{figure}

To summarize, the concepts introduced in this section form a powerful toolbox that we use in the
following to investigate the properties of the Oslo model. The first of these tools are
the evolution operators, that generate the evolution trees starting from a configuration.
The paths in these trees correspond to particular sequences of configurations and instructions,
the latter being represented as Stacks of Instructions (\SI{}, \SRI{} and \STI{}).
The information about the order according to which the instructions are performed 
is mathematically encoded in sequences of odometers (sitewise instruction counters).
Finally, a legal \SI{} and sequence of odometers fully determine a unique path
from the initial configuration to the final one.
\section{Classification of recurrent configurations}
\label{sect:classes recur}

In this section, we exhibit an equivalence relation on $\recur$ 
and we investigate the properties shared by the elements in the equivalence classes. 
A first investigation of the distribution between recurrent and transient configurations was proposed
by \cite{chua_exact_2002,dhar_steady_2004}, it was based on the representations we have
introduced in the previous section.
By using the $g$-representation, which combines the advantages of
$h$- and $z$-representations, the equivalence relation appears in a clear way.

\subsection{Elementary properties of recurrent configurations \label{subsect:preliminaries rec conf}}
Recurrent configurations are distinct from transient configurations
from the absence or presence of \emph{forbidden subconfigurations}.
Chua and Cristensen conjectured that two simple conditions were enough to tell a recurrent
configuration from a transient one \cite{chua_exact_2002}. 
Dhar later proved this conjecture \cite{dhar_steady_2004} that
allows to determine the  exact number of recurrent configuration for a fixed size~$L$.

\begin{prop}
\label{prop:0 at L}
In the $z$ representation of a recurrent configuration, one has
\[ z(L)\geq1. \]
\end{prop}

\begin{proof}
By contradiction, suppose we have a recurrent configuration with $z(L)=0$ and consider the
last instruction that happened on site~$L$. It can only be \instr{q}.
But after a \instr{q} instruction at $x=L$, there is a waiting particle at $x=L$, 
so this cannot be the last instruction.
\end{proof}

\begin{prop}[Isolated zeros]
\label{prop:isolated zeros}
In the $z$-representation of a recurrent configuration, a~zero is always isolated, \textit{i.e.}
subconfigurations of the form $\zcfg{\dots00\dots}$ never appear. 
\end{prop}

\begin{proof}
By contradiction, consider two sites $x$ and $x+1$ such that $z(x)=z(x+1)=0$. 
The last instructions at $x$ and $x+1$ are necessarily \instr{q}. 
But after an instruction~\instr{q} at site~$x$, there is a waiting particle at site $x+1$,
and symmetrically there is a waiting particle at site $x$ after an instruction \instr{q} at $x+1$
so, either way, \instr{q} cannot be the last instruction at $x$ and $x+1$ simultaneously.  
\end{proof}

\begin{prop}[Left of zeros]
\label{prop:2 left of 0}
Consider the $z$-representation of a recurrent configuration $r$, and suppose there is a site~$x_0$ where $z(x_0)=0$.
Then $r$ is of one of the forms
\begin{itemize} 
\item $\zcfg{\dots 21_{k}0\dots}$ with $0\leq k\leq x_0-2$;
\item  or $\zcfg{1_{x_0-1}0\dots}$. 
\end{itemize}

\end{prop}

\begin{proof}
Consider the site~$x_0$ where $z(x_0)=0$. Then by Proposition \ref{prop:0 at L}, we have $x_0<L$. 
The last instruction at site~$x_0$ is necessarily \instr{q} since
$z(x_0)=0$. We now consider the instructions that have happened after this instruction on other sites,
there are necessarily at least two such instructions.
A waiting particle has been added at $x_0-1$. Note that the particle
added at $x_0+1$ will not affect the evolution of the left part, $x<x_0$ of the system, because
no further instructions is executed at~$x_0$. 
The last instruction at $x_0-1$ cannot be \instr{t}, because it would
send a particle to~$x_0$, it can also not be~\instr{q} by
Proposition.~\ref{prop:isolated zeros}. So it is either \instr{p} or
\instr{s}. In the first case (instruction \instr{p}), 
we have $z(x_0-1)=2$ and this corresponds to the first case of the Proposition.
In the second case (instruction \instr{s}), 
we have $z(x_0-1)=1$ and it was necessarily preceded by an instruction
\instr{q} at site~$x_0-1$. 

Thus, we have either $z(x_0-1)=2$ (first case) or $z(x_0-1)=1$ and
a similar situation (the last instruction is \instr{q}) at $x_0-1$. Applying
repeatedly the same reasoning either ends when a~$x'<x_0$ is such that $z(x')=2$
(first case) or when it reaches the site $x=1$ with $z(1)=1$ (second case). 
\end{proof}

\begin{prop}[Right of zeros]
\label{prop:2 right of 0}
Consider the $z$-representation of a recurrent configuration $r$. 
Suppose there is a site~$x$ where $z(x)=0$, then $r$ equals 
$\zcfg{\dots01_{k}2\dots}$ for some $0\leq k\leq L-x-1$
\end{prop}

\begin{proof}
The proof is similar to the preceding one (Proposition \ref{prop:2 left of 0}) with a
symmetrical process to the right.
It remains to consider the case where $z(L-1)=0$ and show that we necessarily have $z(L)=2$.
By contradiction, suppose that we have $z(L-1)=0$ and $z(L)=1$. 
Then before the last instruction \instr{q} at $L-1$, we have a 
configuration  of the form $a_{L-1}\zcfg{\dots10}$.
If this is a valid situation, then an instruction \instr{p} could occur instead of 
\instr{q} at $L-1$, but this would end the stabilization in the configuration~$\zcfg{\dots20}$.
This situation is however not possible because, from Proposition~\ref{prop:0 at L}, 
no recurrent configuration have $z(L)=0$.
$z(L)$ must therefore be equal to $2$.
\end{proof}

\begin{theorem}[Chua and Christensen 2002 and Dhar 2004]
\label{chuachris}
Consider a stable configuration~$c\in\stable$, then $c\in\recur$ if and only if it
respects the conditions of Propositions~\ref{prop:2 left of 0} and~\ref{prop:2 right of 0}.
\end{theorem}
We have already shown that if $c\in\recur$, 
the Propositions
\ref{prop:2 left of 0} and~\ref{prop:2 right of 0} are necessarily verified.
This theorem shows that they are sufficient. It was conjectured by Chua and Christensen and proven by Dhar
with a modified toppling matrix \cite{dhar_steady_2004}. 
Furthermore, 
Chua and Christensen showed that \cite{chua_exact_2002}
\[ \big|\recur(L)\big|\leq \tfrac{\varphi^{2L+1}+\varphi^{-(2L+1)}}{\varphi+\varphi^{-1}}
    = F_{2L} \]
where $\varphi=\tfrac{1+\sqrt5}2$ is the golden ratio and $F_n$ is the $n^{\text{th}}$ Fibonacci
number ($F_0=F_1=1$, and $F_{n+2}=F_n+F_{n+1}$). 

The Propositions \ref{prop:2 left of 0} and \ref{prop:2 right of 0} validate the use
of the $g$-representation for configurations, because it follows from them
that we have $g_r(x)\geq 0$ for any $r\in\recur$.
The distinction into ramparts, in particular, reflects the alternating rule for $0$s and $2$s
(see Figure \ref{fig: crenels merlons}).
The first application of the $g$-representation is to provide a straightforward proof of
Theorem~\ref{chuachris}.

\begin{proof}
In the $g$-representation, it is actually easy to see that
any configuration respecting Propositions~\ref{prop:2 left of 0} and~\ref{prop:2 right of 0}
is accessible from the initial configuration~$a_1\zcfg{2_L}$ following a sequence of legal instructions.
To do so, consider the avalanche of which the associated \SRI{} are made
of $\instr{q}$ for all toppling instructions inside $\domain[r]$ and $\instr{p}$ in $\final[r]$.
Now, perform the toppling instructions and dissipate stones rampart by rampart, from the upmost to the lowest 
and from left to right. This shows that the conditions
\ref{prop:2 left of 0} and \ref{prop:2 right of 0} are sufficient which finishes the proof.
\end{proof}

\subsection{A path invariant~: the branching power}

Let us consider an evolution tree starting from $c_1$ and
all paths starting from $c_1$ to a node with configuration~$c_2$. 
As~$c_2$ can have multiple occurrences in the tree, there are as many different
paths starting from~$c_1$ and ending on~$c_2$.
The probability of each of these paths is, as explained in 
the Section~\ref{sec:paths}, a product of powers of $p$ and $q$
\[ \mathbb{P}(k)=p^{m_k} q^{n_k} \]
where $k$ denotes one path from~$c_1$ to~$c_2$. 
Each factor $p$ or $q$ corresponds to a branch split in the evolution tree,
therefore the number~$m_k+n_k$ is the number of branch splits that the
path~$k$ has visited, it is called the \emph{branching power} of the path.

\begin{prop}[Invariance of the branching power]
\label{prop:inv branching}
Consider two generalized configurations~$c_1$ and $c_2$ related by at least
one path. Then the branching power is independent of the path from $c_1$ to $c_2$.
\end{prop}

\begin{proof}
The generalized configurations~$c_1$ and~$c_2$ define a unique
toppling function~$\toppling[12]$, the number of topplings at 
site~$x$, $\toppling[12](x)$, only depends on $c_1$ and $c_2$.
Let us select a site~$x$ where $\toppling[12](x)\geq1$ and consider the stack at this site.

Let us denote $m_k(x)$ and $n_k(x)$
the number of instructions \instr{p} and \instr{q} performed at site~$x$ along the path~$k$, 
respectively. 
From the Table~\ref{tab:cases}, we observe that random instructions \instr{p} and \instr{q} 
are triggered at a site~$x$ only when $z(x)=1$, and that if an instruction is performed
at $x$ after them, it is necessarily \instr{t} and \instr{s} respectively. 
Therefore, the \SI{} contains pairs \instr{p}\instr{t} and \instr{q}\instr{s},
the number of which is equal to~$\toppling[12](x)$ up to adjustments depending on the
initial and final configurations only. We conclude that the sum $m_k(x)+n_k(x)$ depends
only on~$z_1(x)$, $z_2(x)$ and $\toppling[12](x)$,  but not on the path~$k$. 
\end{proof}

We immediately deduce from this proposition that 
the total probability, the sum of~$\mathbb{P}(k)$ over all paths from 
$c_1$ to~$c_2$ is a polynomial in the variables $p$ and $q$ of homogeneous degree~$\kappa_{12}$.
We call this degree the \emph{branching power} from $c_1$ to $c_2$, determined by any path~$k$:
The total probability $P_{c_1 \to c_2}$ takes the form
\begin{equation}
P_{c_1 \to c_2}=\sum_{\substack{\text{paths }k\\\text{from }c_1\text{ to }c_2}} \mathbb{P}(k) 
      =\sum_{i} \gam{i}\, p^{\alpha_i} \, q^{\kappa_{12}-\alpha_i}.
\label{eq:probability}
\end{equation}
Whenever the configuration $c_1$ is the initial configuration $a_1\zcfg{2_L}$, 
we simply write $P_c$ instead of $P_{a_1\zcfg{2_L} \to c}$. 

\begin{remark}
For given $c_1, c_2\in\general$, 
the coefficients $\gam{i}$ depend on the evolution tree, or equivalently, on the
evolution operator $\evol$. However, if $c_2\in\stable$, 
the Abelian property of the model imposes that the coefficients
$\gam{i}$ do not depend on the evolution tree. 
\end{remark}

\subsection{Lifted configurations}
Let us define the \emph{lifted} configuration of a generalized configuration $c=(z, \,w)$.
The lifted configuration, denoted by $\lift c$, is the generalized configuration 
$\lift c=(\lift z,\lift w)$ where for all~$x$
\begin{equation}
\label{def:lifted}
\lift z(x)=\begin{cases}z(x) & \text{if $z(x)\leq 1$} \\ z(x)-1 & \text{if $z(x)=2$}\end{cases}
\qquad 
\lift w(x)=\begin{cases}w(x) & \text{if $z(x)\leq 1$} \\ w(x)+1 & \text{if $z(x)=2$}\end{cases}
\end{equation}
\begin{prop}\label{prop:lift injective}
The correspondence between stable configurations (for which $w\equiv0$) and their
lifted configurations is one-to-one. 
\end{prop}

\begin{proof}
The unlifting operation consists in perfoming instructions \instr{p} at each site with a
waiting particle. 
It then suffices to prove the injectivity of the lifting operation. 
Consider $c_1, c_2\in\stable$ and their respective lifted configurations
$\lift c_1$ and $\lift c_2$. If $c_1\neq c_2$ then there is a site $x$ where
$z_1(x) > z_2(x)$ or $z_1(x) < z_2(x)$. Let us consider, without loss of generality,
that we are in the first case. If $z_1(x)=2$, then $\lift w_1(x)=1$ and $\lift w_2(x)=0$
and therefore $\lift c_1\neq\lift c_2$. Otherwise, 
$\lift z_1(x)=z_1(x)>z_2(x)=\lift z_2(x)$ and again $\lift c_1\neq\lift c_2$. 
\end{proof}

We focus on the set of recurrent configurations
$\recur\subset\stable$, which is the only set of stable configurations relevant for the stationary state. 

\begin{prop}
\label{prop:lift}
Let us consider a recurrent configuration~$r\in\recur$.
There exists an evolution operator~$\evol$ such that all paths
of the evolution tree starting at $a_1\zcfg{2_L}$ and ending at configuration $r$ contain
the lifted configuration~$\lift r$.
\end{prop}

\begin{proof}
Consider the \SRI{}, denoted by $S$, of a path from $a_1\zcfg{2_L}$ to $r\in \recur$.
By construction, the associated domain $\final[r]$ of $S$ is filled with \instr{p}. 
The $x$ coordinates of these positions give the sites with $2$ particles in $r$.
We also know that these \instr{p} are the last instructions executed for the corresponding stacks.
Locally, any $x$ where $z_r(x) = 2$ must have a waiting unit on site before the last \instr{p} 
instruction was executed.
Using the Ablelian property, one can postpone the execution of instructions of the waiting 
units in $\final[r]$ to the end of the stabilization process, 
since no further instruction is executed at these sites after them.
The configuration where all instructions have been performed except the last \instr{p} 
is by definition~$\lift r$, it is the result of the execution of all the instructions on $\domain[r]$ of~$S$.
\end{proof}

\begin{remark}
The proposition~\ref{prop:lift} associates an evolution operator~$\evol$ to each 
recurrent configuration. One may ask whether this operator could be the same for all the 
recurrent configurations in~$\recur$ or, in other words, if there is a systematic way to 
reach the lifted configurations
(the operator used to build Figure \ref{fig:path tree L=2} is such an operator). 
The answer to this question is positive for system sizes up to $L=3$ but 
is negative for larger system sizes.
\end{remark}

\subsection{Natural configurations}
\begin{definition}[Natural configurations]
The set of natural configurations is defined as follows 
\begin{equation}
\recurnat=\left\{r\in\recur\;, \forall x \in \intrange{1}{L-1}, z_r(x)\geq1 \text{ and }
z_r(L)=2 \right\}.
\label{def:Rnat}
\end{equation}
\end{definition}
Natural configurations are a small subset of $2^{L-1}$ stable configurations and are 
central to our analysis of the stationary state.
In the large $L$ limit, the scaling of its cardinal is exponentially 
smaller than the cardinal of~$\stable$, which is $3^L$, or that of~$\recur$, which is 
$F_{2L}$ (see Theorem~\ref{chuachris}). 

\begin{prop}
\label{prop:natural class}
Consider a recurrent configuration~$r\in\recur$. 
Then there exists at least one evolution operator~$\evol$
and a unique natural configuration $\nat r\in\recurnat$ such that all
paths, produced by $E$, from the configuration~$a_1\zcfg{2_L}$ to~$r$, contain
the lifted configuration~$\liftnat r$,
Moreover, for each of these paths, the path part from $\liftnat r$ to $r$ is 
made of the same sequence of configurations and is, in that sense, unique. 
\end{prop}

\begin{proof}
We proceed as in Proposition~\ref{prop:lift} and build
a path backwards from a lifted configuration.
This path will define a legal order of activation, and so an operator $\evol$ satisfying the statement of the proposition.

From Figure~\ref{fig:stones}, we remark that only instruction \instr{q} can create new crenels.
Note also that crenels created by \instr{q} can start at $x=1$, but only
on the highest rampart.
Let us consider the \SRI{} $S$ (see Figure~\ref{fig:reverse path}) 
of a path from $a_1\zcfg{2_L}$ to a lifted configuration $\lift r$ and let us 
build a backward legal sequence of lifted configurations and odometers $(\lift r_k)_k$ and $(\mu_k)_k$
starting from $\lift r_0=\lift r$ and $\mu_0(x)=\mu_r(x)-\mathbf1_{\final[r]}(x)$~:
Consider the $g$-representation of $\lift{r}_0$ 
and the rightmost site~$x'$ of a crenel~$K$ (that is such that $g(x'+1)>g(x')$),
then the last instruction at site~$x'$ is necessarily a~\instr{q} because a~\instr{t}
would move the stone down one level and \instr{s} and \instr{p} do not move stones.
Not executing this instruction from the corresponding stack of $S$ results in 
a new odometer $\mu_1(x)=\mu_0(x)-\mathbf1_{\{x'\}}(x)$ and an associated
generalized configuration~$\lift{r}_1$ that differs from~$\lift{r}_0$ 
at sites~$x'$ and $x'+1$ on the level of~$K$. Moreover, the stone at site~$x'-1$ is deactivated (waiting to stable), if there is any.
As a conclusion, this backward instruction moves a waiting stone from site~$x'+1$ to site~$x'$.

By repeating this backward \instr{q} move, all crenels from~$\lift{r}$ can be removed (see Figure \ref{fig:reverse path}).
The order in which the movements are performed is not important, in virtue of the Abelianness of the stabilization.
The configuration obtained at the end of the procedure has a single merlon on each rampart, 
and all of these merlons start at~$x=1$.
To obtain a lifted natural configuration, note that it is possible to add extra backward moves and introduce 
waiting stones from the dissipation sink until the lowest rampart contains exactly $L$
stones, the rightmost one of which (at $x=L$) is waiting, making the final configuration of the sequence of movements
a lifted natural configuration~$\liftnat r$. 
Since no stones were moved between ramparts, we see that the configuration~$\liftnat r$ 
is unique and therefore $\nat r$ also.

Finally, we remark that all the instructions that we performed backward
define a unique backward path from $r$ to $\nat r$, or in the forward time evolution, from $\nat r$ to $r$. 
\end{proof}

Thanks to this Proposition, we can introduce the notation $\nat r$ to denote the natural
configuration associated to a $r\in\recur$.

\begin{figure}[ht!]
\centering
\begin{tikzpicture}
    \begin{scope}[yshift = 1cm]
        \reprgg{0/1,0/0,1/0,1/0,0/1,0/0,0/1,0/0,1/0,0/1}
        \node[rectangle, draw] at (-1,0.5) {$\lift r$};
    \end{scope}
  \begin{scope}[xshift = 0 cm]
    \reprgg{1/0,0/1,0/0,1/0,0/1,0/0,0/1,0/0,1/0,0/1}
    \topplingleftarrow{0,0,1,0}{0.5}{blue};
\end{scope}
\begin{scope}[xshift = 0cm, yshift = -1 cm]
    \reprgg{1/0,1/0,0/1,0/0,0/1,0/0,0/1,0/0,1/0,0/1}
    \topplingleftarrow{0,0,0,1}{0.5}{blue}
\end{scope}
\begin{scope}[yshift = -2 cm]
    \reprgg{1/0,1/0,1/0,0/1,0/0,0/0,0/1,0/0,1/0,0/1}
    \topplingleftarrow{0,0,0,0,1}{0.5}{blue}
    \node at (3,-0.4) {$\vdots$};
\end{scope}
\begin{scope}[xshift = 0cm, yshift = -3.5 cm]
    \reprgg{1/0,1/0,1/0,1/0,1/0,1/0,0/1,0/0,0/0,0/0}
\end{scope}
\end{tikzpicture}
\caption{\label{fig:reverse path} Example of a reverse path starting at the lifted configuration
$\lift r$ with $r\in\recur$. All crenels are removed while the stones stay in the same rampart.
Arrows are \instr{q} instructions appearing in 
all paths from $a_1\zcfg{2_L}$ to $r$ that we reversed in legal way.
The head of the arrow points toward the site at which the latest instruction \instr{q} is being canceled.
Remark that such a site corresponds to the rightmost site of one of the crenels.}
\end{figure}

\begin{definition}[Natural equivalence]
\label{def:natural equivalence}
For two configurations~$r_1$, $r_2\in\recur$ we define the \emph{natural equivalence}
$r_1\sim r_2$ by 
\[ r_1\sim r_2 \text{ if and only if }\nat r_1=\nat r_2. \]
\end{definition}
This defines a partition of $\mathcal{R}$ into natural equivalence classes.
We denote $[r]$ the equivalence class of~$r$ in $\recur$. 

\begin{corollary}[Natural factorization]\label{prop:equivalence coeff}
Denote by $P_{c,\,\evol}(p,\,q)$ the $p,q$ polynomials weighing the transition~$a_1\zcfg{2_L}\to c$, which depends explicitly on the choice of $\evol\in\evolution$.
In particular, when there is no doubt on the choice of $\evol$, we simplify the notation $P_{c,\,\evol}(p,\,q) = P_{c}(p,\,q)$. 
For example, if $c\in\recur$, we know that $P_{c,\,\evol}(p,\,q)=P_{c}(p,\,q)$, in virtue of the Abelianness of the stabilisation \cite{dhar_steady_2004}.
Then, for each $r\in[\nat r]$, with $\nat r \in\recurnat$, we can find one $\evol\in\evolution$ such that 
\begin{equation}
P_r(p,\,q)=P_{\liftnat r,\,\evol}(p,\,q)\; p^{\pi_r}\, q^{\theta_r}=P_{\liftnat r}(p,\,q)\; p^{\pi_r}\, q^{\theta_r}
\label{eq:factorization}
\end{equation}
where $\pi_r=|\final[r]|$ and $\theta_r=|\domain[r]|\,-\,|\domain[\liftnat r]|$. 
\end{corollary}

\begin{proof}
From Proposition \ref{prop:natural class}, we can consider a maximal evolution tree such that all paths from~$a_1\zcfg{2_L}$ to the elements of~$[r]$
contain the configuration~$\liftnat r$. 
As we have seen, the instructions to go from $\liftnat r$ to $r$ define a unique path.
In the $g$-representation, these instructions 
shift stones to the right on the same level and can dissipate stones from the
lowest level when they reach the sink.
Keeping the stones on the same level forbids the use of~\instr{t}.
Therefore, all \instr{p} instructions are stabilizing 
and only \instr{q} contribute to the topplings. 
Therefore, one can find $\evol\in\evolution$ such that the probability of $r$ is given by Equation \eqref{eq:factorization}
where $\pi_r$ is the number of sites~$x$ where $z(x)=2$ and 
$\theta_r$ is the number of topplings at a constant level necessary to perform the path $\liftnat r\to r$. 
The dependency upon the evolution operator can be made implicit, as we only look upon $\evol\in\evolution$ satisfying the above desirable factorization property.
\end{proof}

This result implies that the stationary state, defined by Equation \eqref{eq:stst relation}, factorises as
\begin{equation}
    \psi = \sum_{\nat r\in\recurnat}P_{\liftnat r}(p,\,q) 
    \bigg(\sum_{r\in [\nat r]}p^{\pi_r}q^{\theta_r}\zcfg{r} \bigg)
\end{equation}
It is a direct consequence of the definition of equivalence classes. 
We provide the listing of the $P_{\liftnat r}$ polynomials for $L=3$ in Table \ref{tab:classes L3}.

\subsection{Invariants}

\begin{prop}[Range of the polynomial exponents]
\label{prop:alpha max}
The probability of any configuration
$r\in\recur$ in the stationary state 
takes the form
\begin{equation}
P_r(p,\,q)=
  \sum_i \gam{i} \; p^{\alpha_i} \, q^{\kappa_r-\alpha_i}
   \text{ with } 
    \min_i\alpha_i = \kappa_r - \tau_r + L 
\label{eq:alpha min}   
\end{equation}   
where $\kappa_r$ is the branching invariant and $\tau_r$ the total number of topplings from $a_1\zcfg{2_L}$ to $r$.
\end{prop}

\begin{proof}
Consider a path from $a_1\zcfg{2_L}$ to $r\in\recur$ in a maximal evolution tree. 
The probability weight of this path is completely encoded in the associated \SRI{} denoted $S$. 
In particular, we know that topplings in $S$ are described inside the domain $\domain[r]$ and are either 
\instr{q} or colored by a \instr{p} instruction, that we know to be followed by a \instr{t} 
in the corresponding \SI{}.
One can take the path for which the toppling domain contains only \instr{q}.
The number of \instr{q} instructions is obviously maximal in that case so that we have
\begin{equation*}
  \max_i (\kappa_r-\alpha_i) = \sum_{x=1}^{L}\big(\toppling(x)-1\big) = \tau_r - L.
\end{equation*}
where the $-L$ comes from the deterministic topplings \instr{t} that dissipate $a_1$ starting from $a_1\zcfg{2_L}$.
\end{proof}

\begin{corollary}[Branching power]
The branching power~$\kappa_r$ of a recurrent configuration~$r$ is
\begin{equation}
 \kappa_r = \pi_r + \tau_r - L.
\label{eq:kappa}
\end{equation}
\end{corollary}

\begin{proof}
The corollary follows from Equation~\eqref{eq:alpha min} with $\min_i\alpha_i=|\final[r]|=\pi_r$.
\end{proof}

\begin{prop}\label{prop:delta class inv}
For $r\in\recur$ and $y\geq 0$, we define the quantity 
\[ \sigma_r(y) = L+1-y -\big|\{x, g_r(x)\geq y\}\big|. \]
where $g_r(x)$ is the $g$-representation of $r$.
It is the number of stones dissipated from height (or the level) $y$ in any path from $a_1\zcfg{2_L}$ to $r$. 
Then, the quantity 
\begin{equation}\label{eq: delta}
\delta_r  = \sum_{y=2}^L (y-1) \; \sigma_r(y)
\end{equation}
is a path invariant, for fixed $r$, counting the number of downward movements of stones 
(the contribution of $a_1$ from $a_1\zcfg{2_L}$ taken aside).
It is also a class invariant, i.e. a constant for any $r'\in[r]$.
Remark that dissipative steps, i.e. topplings at $x=L$, are not considered as downwards movement.
\end{prop}

\begin{figure}
\begin{center}
  \begin{tikzpicture}
    \begin{scope}
       \xaxis{3.3cm}{1,...,5}
       \yaxis{3cm}{1,2,3,4,5}{}
       \node[rectangle,draw] at (1.7,2.5) {$c$};
       \reprgg[black]{5/0,4/0,3/0,2/0,1/0}
       \node[red] at (0.25,-0.25+0.5*5){4};
       \node[orange] at (0.25,-0.25+0.5*4){3};
       \node[violet] at (0.75,-0.25+0.5*4){3};
       \node[blue] at (1.25,-0.25+0.5*3){2};
   \end{scope}
   \draw[thick,->] (3,1) -- (4.5,1);
   \begin{scope}[xshift = 5cm]
       \xaxis{3.3cm}{1,...,5}
       \yaxis{3cm}{1,2,3,4,5}{}
       \node[rectangle,draw] at (1.7,2.5) {$c'$};
       \reprgg{3/0,3/0,2/0,2/0,1/0} 
   \end{scope}
\end{tikzpicture}
\caption{\label{fig:invariant} 
An example of evolution from the generalized
configuration~$c$ to the generalized configuration $c'$ in $g$-representation and 
without specifying the stable and waiting stones. 
Such an evolution involves necessarily $4+3+3+2$ steps where a stone moves down one level. 
Dissipation processes do not count as downward steps.}
\end{center}
\end{figure}

\begin{proof}
Consider $r\in\recur$ and its $g$-representation $g_r(x)$.
The maximal configuration in~$g$-re\-pre\-sen\-ta\-tion has $L+1-y$ stones at the rampart with height $y$. 
The number $\big|\{x, g_r(x)\geq y\}\big|$ is the number of stones at height $y$ in the configuration~$r$
so $\sigma_r(y)$ is the number of stones removed from the level~$y$ from the
maximal configuration~$\zcfg{2_L}$.
These stones have been dissipated, they must have performed exactly~$y-1$ downward movements 
plus one dissipation step, independently of the path.
Multiplying the number of such missing stones from heights $y=2$ to $y=L$ by the number of 
downward steps $y-1$ and summing over all heights is therefore path invariant. 
The choice of $y=2$ instead of $y=1$ in Equation \eqref{eq: delta} takes into account the fact that
dissipation steps are not downward movements, and make the quantity invariant inside a class.
Indeed, any $r\sim r'$ must have the same number of stones for all heights $y>1$ only,
this can be seen from the proof of Proposition \ref{prop:natural class}.
\end{proof}

\begin{prop}\label{prop:nb config}
The probability of~$r$ in the stationary state, 
expressed as in Equation~\eqref{eq:probability}, can be recast into 
\begin{equation}\label{eq:decomp NESS prob}
  P_r(p,\,q) = p^{\pi_r} q^{\nu_r} 
      \sum_{i=0}^{\delta_r} \gam[r]{i} \, p^{i}\, q^{\delta_r-i} 
\end{equation}
where $\nu_r=\tau_r-L-\delta_r=\kappa_r-\pi_r-\delta_r$ and $\gam[r]{i}>0$ for all $i$. 
\end{prop}

\begin{proof}
The total number of topplings along a path from~$a_1\zcfg{2_L}$ to~$r$ is~$\tau_r$.
The number of topplings, initial \instr{t} put aside, corresponding to a stone downward movement is $\delta_r$.
Let us call~$\nu_r$ the number of topplings corresponding to stone movements at fixed height and dissipation steps.
We have the relation $\tau_r-L=\delta_r+\nu_r$ as topplings can only be of two types: either between two levels or at constant level (dissipative topplings belong to the latter).
Also, from Equation \eqref{eq:kappa} we have directly the relation $\kappa_r=\nu_r+\delta_r+\pi_r$.
We conclude that from any path from~$a_1\zcfg{2_L}$ to $r$, there is at least~$\nu_r$ instructions~\instr{q} 
and $\pi_r$ instructions~\instr{p}. 

We shall now inspect if we can find a set of paths for which the $\delta_r$ instructions are not constrained,
in the sense that these could be set indifferently to \instr{p} or \instr{q} instructions.
If true, each of these $\delta_r$ instruction would add one to the number of terms in the 
polynomial~$P_{\liftnat r}$,
such that this polynomial has precisely $\delta_r+1$ terms.

As $\delta_r$ is invariant in a class (see Proposition \ref{prop:delta class inv}), we can take $r\in\recurnat$. 
Now, it suffices to consider the following avalanche: start from $a_1\zcfg{2_L}$ and dissipate, 
using only $\instr{q}$ instructions, all the stones one after another.
Do so by always choosing to dissipate the leftmost (waiting) stone (of the intermediate
configuration) which is missing in the final configuration $r$.
Along this procedure, we dissipate level by level, from the top to the bottom and, 
for fixed height, from right to left the stones.
When all the necessary stones are dissipated, we just stabilize the remaining waiting units with
$\instr{p}$ and we obtain the desired $r$ configuration.
Notice that along the procedure, \emph{all the $y-1$ downwards movements} necessary to dissipate a given 
stone at height $y$ happen sequentially at the right of the system where a cluster of $y$ waiting units exists
(see Figure \ref{fig:order dissip delta}).
This must be true as we enforced the dissipation to be performed with stones level by level.
Now, to unstabilize a cluster of $y$ 
adjacent waiting units, we need at least one \instr{q} instruction executed, 
and all the other waiting units can execute \instr{p} or \instr{q} indifferently. 
So, for each stone at the level~$y$, we can have a number in $\llbracket 0,y-1 \rrbracket$ of \instr{p} 
instructions compatible with the dissipation process.
The upper bound $y-1$ is maximal, otherwise the stone cannot be dissipated and 
must stop before reaching the sink.
Adding up the contribution from all dissipated stones
entails relation \eqref{eq:decomp NESS prob}.
\end{proof}

\begin{figure}[ht!]
\begin{center}
  \begin{tikzpicture}
    \begin{scope}
       \xaxis{3.3cm}{1,...,5}
       \yaxis{3cm}{1,2,3,4,5}{}
       \reprgg[black]{5/0,4/0,3/0,2/0,1/0}
       \node[red] at (0.25,-0.25+0.5*5){1};
       \node[orange] at (0.25,-0.25+0.5*4){3};
       \node[violet] at (0.75,-0.25+0.5*4){2};
       \node[blue] at (1.25,-0.25+0.5*3){4};
   \end{scope}
   \begin{scope}[xshift = 6cm]
       \xaxis{3.3cm}{1,...,5}
       \yaxis{3cm}{1,2,3,4,5}{}
       \reprgg[black]{3/0,3/1,2/1,1/1,0/1}
   \end{scope}
\end{tikzpicture}
\caption{\label{fig:order dissip delta}(Left) Same example as in Figure \ref{fig:invariant}
where the order of dissipation is specified for each missing stone of the final configuration. 
(Right) $g$-representation of $a_2a_3a_4a_5\zcfg{02211}$ demonstrating that downward topplings can be set to occur one after another inside a cluster of waiting stones located at the right of the system.
It is clear that we only need one instruction \instr{q} executed among the four waiting stones (gray squares) to make the cluster unstable and dissipate the upmost stone.}
\end{center}
\end{figure}

\subsection{Application for a small system}
In the case $L=3$, the $N_3=13$ recurrent configurations are divided into $2^{3-1}=4$ natural classes.
Table \ref{tab:R3} encodes partially the structure of~$\recur(3)$ with respect to the invariants.
The only remaining problem concerns the evaluation of the $(\gam{i})_i$ sequences.
They are given in Table \ref{tab:classes L3} for $\recur(3)$.
These tables can be build for larger~$L$, but it becomes lengthy to report them.

\begin{table}[ht!]
\renewcommand\arraystretch{2.5}
\centering
\caption{\label{tab:classes L3}Natural classes of the 3 sites Oslo model.
The polynomial $P_{\liftnat r}$ is class invariant.
The largest class (cardinal $2^L$) is the class containing $\zcfg{1_L}$.}
\begin{tabular}{|p{2.5cm}p{4.0cm}|c|}
\hline
Natural member & Other members & $P_{\liftnat r}$ \\
\hline
\hline
$\zcfg{222}$ & & $1$ \\ 
\hline
$\zcfg{122}$ & & $3p^2+3pq+q^2$\\ 
\hline
$\zcfg{212}$ & $\zcfg{221}, \zcfg{022}$ & $6p^3+9p^2q+5pq^2+q^3$\\ 
\hline
$\zcfg{112}$ & 
$\zcfg{121}$, $\zcfg{202}$, $\zcfg{012}$, $\zcfg{211}$, $\zcfg{021}$, $\zcfg{102}$, $\zcfg{111}$ & 
$18p^4+33p^3q+24p^2q^2+8pq^3+q^4$ \\
\hline
\end{tabular}
\end{table}

\begin{table}
\centering
\caption{\label{tab:R3}Recurrent configurations of 3-site Oslo model, organized by 
natural classes. 
Are listed the class invariant $\delta$ and path invariants $\kappa$ (branching power), $\tau$ 
(number of topplings 
from $a_1\zcfg{2_L}$) and constant factors $\pi$ (least number of \instr{p} instructions), 
$\nu$ (least number of \instr{q} instructions).
The last column illustrates graphically the associated $g$-representation.}
\begin{tabular}{|c|c|cc|cc|c|}
\hline
Configuration & $\delta$& $\kappa$& $\tau$ & $\pi$ & $\nu$ & $g$-representation \\
\hline
\hline
$\zcfg{222}$ & 0 & 3 & 3 & 3 & 0 & \tikz{\reprg{3,2,1}{0.4}{}}\\
\hline
$\zcfg{122}$ & 2 & 5 & 6 & 2 & 1 & \tikz{\reprg{2,2,1}{0.4}{}}\\
\hline
$\zcfg{212}$ & 3 & 7 & 8 & 2 & 2 & \tikz{\reprg{2,1,1}{0.4}{}}\\

$\zcfg{022}$ &  & 8 & 9 & 2 & 3 & \tikz{\reprg{1,2,1}{0.4}{}}\\
$\zcfg{221}$ &  & 8 & 9 & 2 & 3 & \tikz{\reprg{2,1,0}{0.4}{}}\\
\hline
$\zcfg{112}$ & 4 & 9 & 11 & 1 & 4 &  \tikz{\reprg{1,1,1}{0.4}{}}\\
$\zcfg{121}$ &  & 10 & 12 & 1 & 5 & \tikz{\reprg{1,1,0}{0.4}{}}\\
$\zcfg{202}$ &  & 12 & 13 & 2 & 6 & \tikz{\reprg{1,0,1}{0.4}{}}\\
$\zcfg{012}$ &  & 12 & 14 & 1 & 7 & \tikz{\reprg{0,1,1}{0.4}{}}\\
$\zcfg{211}$ &  & 12 & 14 & 1 & 7 & \tikz{\reprg{1,0,0}{0.4}{}}\\
$\zcfg{021}$ &  & 13 & 15 & 1 & 8 & \tikz{\reprg{0,1,0}{0.4}{}}\\
$\zcfg{102}$ &  & 14 & 16 & 1 & 9 & \tikz{\reprg{0,0,1}{0.4}{}}\\
$\zcfg{111}$ &  & 14 & 17 & 0 & 10 & \tikz{\reprg{0,0,1}{0.4}{white}}\\
\hline
\end{tabular}
\end{table}
\section{Counting the paths}\label{sect:count paths}
The classification of the recurrent configurations reveals the existence of the polynomials
$P_{\liftnat r}(p,q)$ with $\tilde r \in\recurnat$.
The coefficients of these polynomials, denoted by $\gam{i}$, 
are consequently class invariant and are related to the number of possible paths
from $a_1\zcfg{2_L}$ to $\tilde r$. In this section, we investigate the values taken by these coefficients,
and show that they are determined by several constraints on the toppling domain. 
Formulating these constraints as coloring problems, we obtain an expression for $\gam{i}$ as
alternating sums yielded by the inclusion-exclusion principle. 

\subsection{Mapping between paths and colorings}
The toppling domain of $r\in\recur$, (recall Definition \ref{def:toppling domain}) 
is an \emph{invariant} geometrical object of $\Z[2]$ for any legal path between $a_1\zcfg{2_L}$ and $r$.
The model defines rules for coloring these domains.

\begin{definition}[Coloring]
\label{def:color_domain}
Let us consider a domain $V\subset\Z[2]$. A 2-\emph{coloring}, or just coloring, of $V$ is a two-valued
function $S:V \to \{\instr{p},\instr{q}\}$. 
We denote by $\coloring(r)$ the set of all 2-colorings on the domain $\domain[r]\cup\final[r]$.
\end{definition}

Colorings are just another denomination of the Stacks of Random Instructions,
but we found it more suggestive given the combinatorial problem to come.
Any object defined on or with the \SRI{} applies immediately to the colorings (e.g. the odometers).

\begin{prop}\label{prop: partial order}
Toppling domains define a partial order on the recurrent configurations. 
\end{prop}
\begin{proof}
Let us show that $\domain[r]=\domain[r']$ if and only if $r=r'$
from the Proposition~\eqref{prop:toppling} 
and the fact that the initial configuration $a_1\zcfg{2_L}$ is fixed. 
Suppose we have two recurrent configurations $r$ and $r'$ such that $\domain[r]=\domain[r']$.
By construction, a toppling at site $x$ makes one particle at $x$ jump towards site $x+1$.
Since the initial configuration is fixed and the toppling domain is defined by the number of random topplings at each site, $r$ and $r'$ must have the same number of stones at each site.
Because $r$ and $r'$ are stable, the only instructions one can perform further
are settlings of the waiting stones from which we conclude that $r$ must equal to $r'$.
We can therefore define a partial order $\leq$ on~$\recur$, induced by the inclusion relation,
where, for $r$ and $r'$ in $\recur$. 
\[ r \leq r' \quad \text{if and only if}\quad \domain[r] \subset \domain[r']. \]
The order is partial because one cannot compare two elements $r,r'\in\recur$ 
when $|\domain[r]\cap\domain[r']|< \min(|\domain[r]|,|\domain[r']|)$.
\end{proof}

As we show in Proposition \ref{prop:non-legal coloring}, this partial order is intimately 
related to the way colorings can or cannot be legal for a given configuration.
\begin{prop}[Maximal toppling domain]\label{prop:max domain}
The toppling domain associated to $\zcfg{1_L}$, which we denote $\domain[\max]$, is maximal in the sense that for any $r\in\recur$,
we have $\domain[r]\cup\final[r]\subseteq \domain[\max]$.
\end{prop}

\begin{proof}
Notice that any \SRI{} of a legal path between $a_1\zcfg{2_L}$ and $\zcfg{1_L}$ is defined on $\domain[\max]\cup\final[\max]=\domain[\max]$.
Indeed, one has $\final[\max]=\emptyset$ because $\zcfg{1_L}$ is the unique recurrent configuration that contains
no stones.
One can see that there always exists a path $\lift r \to \zcfg{1_L}$ for any $r\in\recur$.
This immediately implies $\domain[r]\subseteq\domain[\max]$.
Now, suppose there exists $r\in\recur$ such that we can find one site $x\in\intrange{1}{L}$ for which $|\domain[r]\cup\final[r]|(x)>|\domain[\max]|(x)$. 
From what we have discussed before, it can only be so if $|\domain[r]|(x)=|\domain[\max]|(x)$ and $|\final[r]|(x)=1$.
The fact that $|\final[r]|(x)=1$ implies that a stone has been stabilised at $x$.
This can not be true as $\zcfg{1_L}$ contains no stones, in particular on $\intrange{1}{x}$ and, 
in virtue of $|\domain[r]|(x)=|\domain[\max]|(x)$, so does $r$ at $x$.
Remark that the hypothesis $|\domain[r]\cup\final[r]|(x)>|\domain[\max]|(x)$
contradicts Theorem~\ref{chuachris}, which is enough to conclude.
\end{proof}

Conversely, there is a \emph{minimal toppling domain} $\domain[\min]$ which obviously 
corresponds to the transition to $\zcfg{2_L}$, the first configuration reached from $a_1\zcfg{2_L}$.

The existence of the maximal toppling domain invites to consider not only all paths, but directly all the possible 
colorings on $\domain[\max]$.
We will proof this intuitive result in the Proposition \ref{prop:1 to 1 col and legal path}.
For the moment we do not know 
if a coloring $S$ leads to a legal path, and, if so, to which $r\in\recur$ the generated stabilizing odometer leads to.
We derive hereafter a necessary, yet not sufficient, condition in this direction. 

\begin{prop}\label{prop:final <-> r}
    The initial configuration $a_1\zcfg{2_L}$ and final domain $\final[r]$ 
    define uniquely the recurrent configuration $r$. 
\end{prop}

\begin{proof}
    By construction, $\final[r]=\big\{(x_1,\, t_1),\; (x_2,\,t_2),\; \dots (x_k,\,t_k)\big\}$ 
    is a set of points of $\Z[2]$. 
    For each element~$i$ 
    \begin{itemize}
        \item $x_i$ is a site where $z(x_i)=2$;
        \item $t_i$ is the number of stones dissipated over the 
        subconfiguration on the sites $\llbracket 1,x_i \rrbracket$.
        It is equal to the number of toppling instructions \instr{q} or \instr{t} performed 
        on the site~$x_i$. 
    \end{itemize}
    Moreover, recurrent configurations are correlated so that between 
    $x_i$ and $x_{i+1}$ there is at most one site $x'$ such that $z(x')=0$.
    All other sites in the interval contain $1$ particle.
    
    By contradiction, suppose that there are two recurrent configurations $r_1$ and $r_2$ compatible with $\final[r]$.
    In particular, the $g$-representations of $r_1$ and $r_2$ differ on a range $\intrange{x_i}{x_{i+1}}$ for 
    $i=0,...,k-1$ ($x_0=1)$. 
    A range $\intrange{x_i}{x_{i+1}}$ describes in $g$-representation:
    \begin{itemize}
        \item[a)] either a crenel between two merlons;
        \item[b)] or a transition between the rightmost merlon of the rampart above, 
        and the leftmost merlon of the rampart just below.
    \end{itemize} 
    These two cases are illustrated on the Figure~\ref{fig:proof final domain}.
    \begin{figure}[ht]
    \centering
    \begin{tikzpicture}
    \begin{scope}
    \reprgg{0/1,0/0,0/0,1/0,1/0,1/0,1/0,0/0}
    \node at (-0.5,0.25) {a)};
    \node at (0.25,-0.25) {$x_i$};
    \node at (3.25,-0.25) {$x_{i+1}$};
    \end{scope}
    \begin{scope}[xshift=7cm]
    \reprgg{1/1,1/0,1/0,1/0,1/0,1/0,0/0}
    \node at (-0.5,0.25) {b)};
    \node at (0.25,-0.25) {$x_i$};
    \node at (2.75,-0.25) {$x_{i+1}$};
    \end{scope}
    \end{tikzpicture}
    \caption{\label{fig:proof final domain} The two possible cases of the configuration in $g$-representation
         along a segment out of the support of~$\final[r]$. a) A site with $z=0$ exists between $x_i$ and $x_{i+1}$; 
         b) No such site exists, then all sites have $z=1$.}
    \end{figure}
    
    Since the number of stones dissipated from $I=\intrange{x_i}{x_{i+1}}$
    is equal to $t_{i+1}-t_i$ it cannot differ between $r_1$ and $r_2$.  
    Modifying the $g$-configuration from $r_1$ to $r_2$ is therefore only possible by moving stones 
    between $x_i$ and $x_{i+1}$
    but no transformation of this kind is allowed because moving a stone would add a site with $z=2$,
    either in an existing crenel (case a)) or inside a merlon (case b)).
    Since adding such a site is equivalent to modifying $\final[r]$, no distinct configurations share the same
    $\final$ domain. 
\end{proof}

\begin{definition}[Legal coloring]

A legal coloring for $r\in\recur$ is non-legal for all the other recurrent configurations
(see Propositions \ref{prop:unique stacks} and \ref{prop:SI defines path}).
A legal coloring for $r\in\recur$ is defined as a legal stack of random instructions (see Definition \ref{def:path legal stack}) which, starting from $a_1\zcfg{2_L}$, stabilizes to $r$.
We denote the subset of legal colorings for $r$ by $\colorleg(r)$, where the star stands for ``legal''.
\end{definition}

A legal coloring for $r\in\recur$ is non-legal for all the other recurrent configurations
(see Propositions \ref{prop:unique stacks} and \ref{prop:SI defines path}).
Moreover, from Proposition \ref{prop:final <-> r} and Proposition~\ref{prop:inv final}
we obtain that for any $S\in\colorleg(r)$ we have the property
\[ \forall v\in\final[r],\; S(v)=\instr{p}. \]

\begin{prop}\label{prop:all p matters}
Let us consider a system of size $L$, and $x,\,y$ such that 
$1\leq x\leq L$ and $1\leq y\leq |\domain[\max]|(x):=\toppling[\max](x)-1$. Then there exists at least one $r\in\recur$ 
with slope $z_r(x)=2$ and with a stabilizing odometer $\mu_r$ such that $\mu_r(x)=y$.
\end{prop}

\begin{proof}
We prove this by constructing a sequence of random instructions leading to a stable configuration satisfying the conditions,
reusing a similar construction as in our proof of Theorem~\ref{chuachris} and Proposition~\ref{prop:nb config}. A schematic view of the sequence 
of instructions is displayed in the Figure~\ref{fig:example p all matters}. Along the process we generate a legal sequence of configurations $(c_t)_t$, where $t$ is the number of dissipated stones with an exception only for the transition from the penultimate to the final element,
and the corresponding odometers~$(\mu_t)_t$.
We start our reasoning with configuration $c_0=\zcfg{\w1_L}=\lift{\zcfg{2_L}}$, which is obtained from the
maximal configuration $a_1\zcfg{2_L}$ after a sequence of $L$ instructions~\instr{t}. 
We therefore have a waiting stone at each site.
We also have~$\forall x', \mu_0(x')=0$.
Then, we dissipate waiting stones from left to right, always choosing the leftmost one. Dissipating the
waiting stone from $x'$ requires $L-x'+1$ toppling instructions \instr{q} at all sites between~$x'$ and~$L$.
After each stone is dissipated, the generalized configuration of the system is that of a lifted stable configuration,
and the process could be stopped by performing \instr{p} instruction wherever there is a waiting stone. 
At site~$x$, the odometer increases by~1 whenever the \instr{q} instruction is performed at~$x$, it is therefore equal to
the number of stones dissipated from sites~$x'\leq x$. 
As soon as $\mu_t(x)$ has reached the value~$y-1$, we continue to dissipate waiting stones, until the next random instruction at $x$ has to be assigned, meaning a waiting stone is sitting at $x$.
At this point, settling instructions (\instr{p}) are performed wherever there is a waiting stone in the configuration, and we obtain a recurrent configuration $c_{t_{\max}}\in\recur$ with the desired properties.
\end{proof}

Using this algorithm, it is possible to dissipate all waiting stones from the configuration $\zcfg{\w1_L}$ to reach 
$\zcfg{1_L}$, the odometer of which is $\mu(x)=\toppling[\max](x)-1$. 

\begin{figure}[htbp]
\centering
\begin{tikzpicture}
    \begin{scope}
        \xaxis{3cm}{1,...,5}
        \yaxis{3cm}{1,...,5}{}
        \node at (2,2) {$c_0$};
        \reprgg[black]{4/1,3/1,2/1,1/1,0/1}
        \draw[->,>=latex, black, line width=0.75mm] (2.5,1.5) -- (4,1.5);
    \end{scope}
    \begin{scope}[xshift=4.5cm]
        \xaxis{3cm}{1,...,5}
        \yaxis{3cm}{1,...,5}{}
        \node at (2,2) {$c_1$};
        \reprgg[black]{4/0,3/1,2/1,1/1,0/1}
        \draw[->,>=latex, black, line width=0.75mm] (2.5,1.5) -- (4,1.5);
    \end{scope}
    \begin{scope}[xshift=9cm]
        \xaxis{3cm}{1,...,5}
        \yaxis{3cm}{1,...,5}{}
        \node at (2,2) {$c_2$};
        \reprgg[black]{3/1,3/0,2/1,1/1,0/1}
        \draw[->,>=latex, black, line width=0.75mm] (1.5,-0.5) |-(1.5,-1) |- (-7.5,-1) -- (-7.5,-1.75);
    \end{scope}
    \begin{scope}[yshift = -5cm]
        \xaxis{3cm}{1,...,5}
        \yaxis{3cm}{1,...,5}{}
        \node at (2,2) {$c_3$};
        \reprgg[black]{3/0,3/0,2/1,1/1,0/1}
        \node at (3.5,1.5) {\large \textbf{...}};
    \end{scope}
    \begin{scope}[xshift=4.5cm, yshift = -5cm]
        \xaxis{3cm}{1,...,5}
        \yaxis{3cm}{1,...,5}{}
        \node at (2,2) {$c_{t_{\max}-1}$};
        \reprgg[black]{2/0,1/1,1/0,1/0,0/1}
        \draw[->,>=latex, black, line width=0.75mm] (2.5,1.5) -- (4,1.5);
    \end{scope}
    \begin{scope}[xshift=9cm, yshift=-5cm]
        \begin{scope}
       \xaxis{3cm}{1,...,5}
       \yaxis{3cm}{1,2,3,4,5}{}
       \node at (2,2) {$c_{t_{\max}}$};
       \reprgg[black]{2/0,1/0,2/0,1/0,1/0}
   \end{scope}
    \end{scope}
\end{tikzpicture}
\caption{\label{fig:example p all matters} A sequence of configurations, in $g$-representation, obtained with legal movements, ending in a recurrent configuration $c_{t_{\max}}$ for which the associated odometer satisfies $\mu(3)=8$ and $z_{c_{t_{\max}}}(3)=2$.    
$\mu(3)-1$ corresponds to the number of stones dissipated from sites $1,2,3$.
Stable and waiting stones are respectively the white and grey squares.}
\end{figure}

Proposition \ref{prop:all p matters} means that all positions in the maximal toppling domain 
$\domain[\max]$ can be controlled through the final domains of the recurrent configurations.

\begin{prop}\label{prop:r reached last}
Consider $r\in\recur$, its stabilizing odometer $\mu_r$ and a coloring 
$S\in\coloring(r)$ where \emph{at least} $\final[r]$ is colored with $\instr{p}$. 
Then $S$ generates a stabilizing odometer $\mu$ which verifies
\[ \forall x\in\intrange{1}{L}, \quad \mu(x)\leq \mu_r(x). \]
\end{prop}

\begin{proof}
Let us consider the coloring $S_0$ on $\domain[\max]$ and configuration $r\in\recur$ such that 
\begin{equation*}
    S_0(v) = \begin{cases}
    \instr{p} & \text{if} ~v\in \final[r] \\
    \instr{q} & \text{otherwise.}
\end{cases}
\end{equation*}
By construction $S_0$ is legal for $r$.

Now, we want to modify the coloring at one position $(x_1,\,t_1)\in\domain[r]$ 
so that $S_1(x',\,t')=S_0(x',\,t')$ for all $(x',\,t')\neq(x_1,\,t_1)$ and $S_1(x_1,\,t_1)=\instr{p}$.
We show that $\mu_1(x)\leq \mu_0(x)$ for all $x$.

After executing the instruction at $(x_1,t_1)$ we have a configuration with $z(x_1)=2$ and two possible evolution from there: either site $x_1$ does not receive any waiting unit until it is globally stabilized, meaning $t_1$ is the last instruction at $x_1$ and by construction $\mu_1(x_1):=t_1<\mu_0(x_1)$; or $x_1$ receives at a latter time one waiting unit from a neighbor site which makes $x_1$ deterministically topple (\instr t instruction), leading necessarily to $\mu_1(x) =\mu_0(x)$ for all $x$.
Indeed, in the latter case, the toppling at $x_1$
\emph{only depends on waiting units different from 
the one stopped by the instruction at} $(x_1,\,t_1)$. 
If site $x_1$ receives a waiting unit after the execution of the instruction $S_1(x_1,\,t_1)$, then we can find some
$\evol\in\evolution$ that generates on $S_0$ and $S_1$ exactly the same legal sequence of odometers $(\mu_t)_t$.
It suffices for that to keep, as long as possible, the waiting unit that should normally execute the $t_1$-th random instruction idle 
on site $x_1$.
Doing so, with the hypothesis we necessarily have, at some point, two waiting units on site $x_1$ and then, just after executing $(x_1,\,t_1)$
in both $S_0$ and $S_1$, we arrive in the same configuration (recalling that we implicitly execute deterministic instructions). 
Obviously, after this point, all the rest of the evolution is the same and leads to $r$.
Iterate use of this argument, considering $S_k$ to be a coloring with $k$ extra $\instr{p}$ 
with respect to $S_0$, leads to the desired result.
\end{proof}

\begin{prop}\label{prop:1 to 1 col and legal path}
    Each coloring $S\in\coloring(\zcfg{1_L})$, \textit{i.e. on $\domain[\max]$}, generates a stabilizing odometer, i.e. defines a legal transition from $a_1\zcfg{2_L}$ to some $r\in\recur$.
\end{prop}

\begin{proof}
    It suffices to take $r=\zcfg{1_L}$ in Proposition \ref{prop:r reached last}, and to use Proposition \ref{prop:max domain} to show that for any $r\in\recur$, one has $\mu_{r}(x)\leq \mu_{\zcfg{1_L}}(x):=|\domain[\zcfg{1_L}]|(x)$ with $\mu_{r}$ the stabilizing odometer for any path $a_1\zcfg{2_L}\to r$.
\end{proof}

\begin{prop}\label{prop:unstab f} 
Consider a coloring $S\in \coloring(\zcfg{1_L})$ reaching $r$ with stabilizing odometer $\mu_{r}$.
Deform $S$ into a new coloring $S'$ (reaching $r'$ with stabilizing odometer $\mu_{r'}$) such that there is 
at least one $(x,\,t)\in\final[r]$ with $S'(x,t)=\instr p$ and $(x,\,t)\in\domain[r']$. 
Then there is a site $x'\neq x$ such that $\mu_{r'}(x')\geq \mu_r(x')$. 
\end{prop} 

\begin{proof}
We consider some $S$, $r$ and $\mu_r$ as in the Proposition.
Let us deform progressively $S$ into $S'$, keeping $S(x,t)=S'(x,t)=\instr{p}$, but such that now $(x,t)\in\final[r']$.
From Proposition \ref{prop:r reached last}, it is clear that making
any modification of the type $S(x',\,t')=\instr p\to S'(x',\,t')=\instr q$ 
for $(x',\,t')\in \domain[r]$, would not lead to $(x,\,t)\in\domain[r']$.
So the modification must necessarily happen on some $(x',\,t')\in \final[r]$ which ensures to have a 
toppling associated to the instruction at $(x',\,t')$ and $\mu_{r'}(x')\geq \mu_r(x')$.
\end{proof}

\begin{prop}\label{prop:non-legal coloring}
Consider the configurations $r,r'\in\recur$ with $r\neq r'$ such that
$\final[r']\cap\domain[r] \neq\emptyset$. 
Define the following coloring $S$ on $\domain[\max]$
\begin{equation*}
    S(v) = \begin{cases}
      \instr{p} & \text{if}~v\in \final[r]\cup (\final[r']\cap\domain[r])\\
      \instr{q} & \text{otherwise.}
    \end{cases}
\end{equation*} 

Then the coloring~$S$ cannot correspond to a legal path from $a_1\zcfg{2_L}$ to $r$. 
\end{prop}

\begin{proof}
In the proof, we use the notation $\{U\}_x := \{x\in\intrange{1}{L}|~(x,y)\in U\}$ for any subset $U\subset \intrange{1}{L}\times \Z$.

From Proposition \ref{prop:r reached last}, we know that at most configuration $r$ is legal 
with respect to $S$.
The stabilizing odometer $\mu$ generated by $S$ must then satisfy $\mu(x)\leq |\domain[r]\cup\final[r]|(x)$. 
We show that this inequality is in fact strict for at least one $x$.
To do so, we proceed with a proof by exhaustion, where all different cases are illustrated on the
Figure \ref{fig:proof c legal r}:

\begin{itemize}
    \item[(A)] Suppose $|\domain[r']\cup\final[r']|(x)\leq |\domain[r]\cup\final[r]|(x)$ for all $x$. 
    Then obviously $\mu(x)=|\domain[r']\cup\final[r']|(x)$ and $S$ is legal for $r'$.

    \item[(B)] We can also have $\final[r']\subset\domain[r]$
    but with some positions $x \in \llbracket 1,L \rrbracket$, $x\notin \{\final[r']\}_x$, such that 
    $|\domain[r']|(x)> |\domain[r]|(x)$.
    In that situation, we can see that 
    $\mu(x)\leq |\domain[r']\cup\final[r']|(x)<|\domain[r]\cup\final[r]|(x)$ for all $x\in \{\final[r']\}_x$.
    Indeed, once waiting units have settle by executing $\instr{p}$ instructions on $\final[r']$, 
    they can't be unsettled without a stabilizing odometer satisfying the condition in 
    Proposition \ref{prop:unstab f}. 
    This would contradict Proposition \ref{prop:r reached last} precisely on at least one position 
    $x\in \llbracket 1,L \rrbracket\setminus  \{\final[r']\}_x$ where $|\domain[r']|(x)> |\domain[r]|(x)$. 
    Therefore $S$ cannot be legal for $r$.
    \item[(C)] Finally we treat all cases where $0<|\final[r']\cap\domain[r]|< |\final[r']|$. 
    Supposing all instructions in $\{\final[r']\cap\domain[r]\}_x$ to be executed, we must satisfy Proposition 
    \ref{prop:unstab f} to unstabilize them and reach $r$. 
    This would, again, contradict Proposition \ref{prop:r reached last}, so $r$ is not reached using $S$.
\end{itemize}
\end{proof}

\begin{figure}[htbp]
\centering
\begin{tikzpicture}
    \begin{scope}[scale=0.8]
        \node at (2,-1){(A)};
        \draw[thick, black!70, fill=black!70, domain=0:4] (0,3) -- (1,3) -- ++(90:0.25) -- ++(180:1) -- cycle;
        \draw[thick, black!70, fill=black!70, domain=0:4] (2.5,4) -- (4,4) -- ++(90:0.25) -- ++(180:1.5) -- cycle;
        \draw[thick, black!70, domain=0:4] (0,0) -- (4,0) -- ++(90:4) -- ++(180:2) -- ++(270:1) -- ++ (180:2) -- cycle; 
        \draw[thick, black!30, fill=black!30, domain=0:4] (0,2) -- (1.5,2) -- ++(90:0.25) -- ++(180:1.5) -- cycle;
        \draw[thick, black!30, fill=black!30, domain=0:4] (3,3) -- (4,3) -- ++(90:0.25) -- ++(180:1) -- cycle;
        \draw[thick, black!30, domain=0:4] (0,0) -- (4,0) -- ++(90:3) -- ++(180:1) -- ++(270:1) -- ++ (180:3) -- cycle; 
        \node (P) at (2,4.5) {$\final[r]$}; 
        \draw[->] (P) edge[out=180, in=90] (0.5,3.3);
        \draw[->] (P) edge[out=0, in=90] (3.5,4.3);
        \node (P') at (2,1.5) {$\final[r']$}; 
        \draw[->] (P') edge[out=180, in=270] (1.25, 1.9);
        \draw[->] (P') edge[out=0, in=270] (3.5, 2.9);
        \node (T) at (4.5,3.5) {$\domain[r]$}; 
        \draw[->] (T) edge[out=180, in=0] (3.5,3.75);
        \node (T') at (4.5,1.5) {$\domain[r']$}; 
        \draw[->] (T') edge[out=180, in=0] (3.5,1.2);
        \draw[line width=0.25mm,, <->]  (0,4.5) -- (0,0) -- (4.5,0) node[right] {site};
        \node[rotate=90] at (-0.5,2) {instructions};
    \end{scope}
    
    \begin{scope}[xshift = 4.5cm, scale=0.8]
        \node at (2,-1){(B)};
        \draw[thick, black!70, fill=black!70, domain=0:4] (0,2) -- (1,2) -- ++(90:0.25) -- ++(180:1) -- cycle;
        \draw[thick, black!70, fill=black!70, domain=0:4] (1.75,2) -- (2,2) -- ++(90:0.25) -- ++(180:0.25) -- cycle;
        \draw[thick, black!70, fill=black!70, domain=0:4] (2.5,3) -- (4,3) -- ++(90:0.25) -- ++(180:1.5) -- cycle;
        \draw[thick, black!70, domain=0:4] (0,0) -- (4,0) -- ++(90:3) -- ++(180:2) -- ++(270:1) -- ++ (180:2) -- cycle; 
        \draw[thick, black!30, fill=black!30, domain=0:4] (0,1.25) -- (1.5,1.25) -- ++(90:0.25) -- ++(180:1.5) -- cycle;
        \draw[thick, black!30, fill=black!30, domain=0:4] (3,2.25) -- (4,2.25) -- ++(90:0.25) -- ++(180:1) -- cycle;
        \draw[thick, black!30, domain=0:4] (0,0) -- (4,0) -- ++(90:2.25) -- ++(180:2.5) -- ++(270:1) -- ++ (180:1.5) -- cycle; 
        \draw[line width=0.25mm,, <->]  (0,4.5) -- (0,0) -- (4.5,0);
    \end{scope}
    
    \begin{scope}[xshift = 9cm, scale=0.8]
        \node at (2,-1){(C)};
        \draw[thick, black!70, fill=black!70, domain=0:4] (0,1.5) -- (1,1.5) -- ++(90:0.25) -- ++(180:1) -- cycle;
        \draw[thick, black!70, fill=black!70, domain=0:4] (2.5,3.75) -- (4,3.75) -- ++(90:0.25) -- ++(180:1.5) -- cycle;
        \draw[thick, black!70, domain=0:4] (0,0) -- (4,0) -- ++(90:3.75) -- ++(180:2) -- ++(270:2.25) -- ++ (180:2) -- cycle;
        \draw[thick, black!30, fill=black!30, domain=0:4] (0,2) -- (1.5,2) -- ++(90:0.25) -- ++(180:1.5) -- cycle;
        \draw[thick, black!30, fill=black!30, domain=0:4] (3,3) -- (4,3) -- ++(90:0.25) -- ++(180:1) -- cycle;
        \draw[thick, black!30, domain=0:4] (0,0) -- (4,0) -- ++(90:3) -- ++(180:1) -- ++(270:1) -- ++ (180:3) -- cycle; 
        \draw[line width=0.25mm,, <->]  (0,4.5) -- (0,0) -- (4.5,0);
    \end{scope}
\end{tikzpicture}
\caption{Sketch of the $3$ different cases for the colorings of Proposition \ref{prop:non-legal coloring}. 
Black rectangles represent $\final[r]$, the area under the black line delimits $\domain[r]$, and the same goes for $\final[r']$ and $\domain[r']$ with the gray color.}
\label{fig:proof c legal r}
\end{figure}

\begin{definition}[Set of constraints]
We denote by
\begin{equation}
\setconstraints[r] = \{\final[r']\cap \domain[r] ,\, r'\in\recur\} \setminus\{\emptyset\}
\end{equation}
the \emph{set of constraints} formed by the intersections of all final domains 
(see Definition \eqref{def:Pr}) with $\domain[r]$.  
The empty set is removed for convenience.
\end{definition}

The latter set of constraints contains all the information we need to separate between a legal and non-legal
coloring for a given $r$.
\begin{lemma}
\label{lemma:color toppling}
Consider~$r\in\recur$ and a coloring~$S$ such that $S(v)=\instr{p}$ for all $v\in\final[r]$ and
\begin{equation}\label{eq:condition legal coloring}
\forall f \in \setconstraints[r], \;
\big|\{v\in f,\, S(v)=\instr{p}\}\big| < |f| 
\end{equation} 
Then $S$ is a legal coloring for $r$.
It allows to give a constructive definition to the set $\colorleg(r)$ of legal colorings to $r$  
\begin{multline}
\label{eq:set legcolor}
\colorleg(r):=\{S\in\coloring(r) \,|~\forall v\in\final[r],~ S(v)=\instr{p}
~\text{and}~\forall f\in\setconstraints[r], \big|\{v\in f, S(v)=\instr{p}\}\big|< |f|\}
\end{multline}
\end{lemma}

\begin{proof}
Using Proposition \ref{prop:non-legal coloring}, the condition \eqref{eq:condition legal coloring} imposes that the path cannot 
stop before at least reaching $r$. 
    On the other hand, Proposition \ref{prop:r reached last} tells us that $r$ is the last configuration 
    that can be reached with the coloring $S$. 
    So $S$ is legal for $r$.
    The domain  $\domain[\max]\setminus(\domain[r]\cup\final[r])$ is not relevant as no instructions 
    there are executed. 
    Also, $\final[r]$ is a constant in all the legal colorings to $r$ and must be colored with \instr{p}.
\end{proof}

\begin{prop}
Taking $r\in \recurnat$, the coefficient $\gam[r]i$
defined in \eqref{eq:probability}, with
$i\in\llbracket 0, \delta_r-1\rrbracket$ equals
\begin{equation}
\label{eq:gamma color}
  \gam[r]i = \Big|\Big\{S\in\colorleg(r), \big|\{v\in\domain[r], S(v)=\instr p\}\big| = i\Big\}\Big|
\end{equation}
\end{prop}
\begin{proof}
  Because of the one-to-one correspondence between paths and colorings, 
  the number of legal colorings is equal to the number of legal paths. 
  By construction, $\gam[r]i$ counts the paths with a factor $p^{\pi_r+i}$, which translates in the colorings
  involving exactly $i$ instructions $\instr{p}$ over $\domain[r]$.
\end{proof}

\subsection{Method of counting}
In this section, we provide a solution to
the problem of the counting and derive an explicit expression for the $\gam{}$ coefficients.
We start by partitioning $\colorleg$ sets with two natural parameters.
\begin{definition}[Composite final domains] 
Given $r\in\recur$, we define the set of \emph{composite final domains} 
made of the union of $k$ final domains and with cardinal $\ell$ as 
\begin{equation}
\compset[r](k,\,\ell):= \Big\{V\subset \domain[r],\; |V|=\ell \; \text{and} \;
  \exists F \subset \setconstraints[r], \; |F|=k, 
  V=\bigcup_{f\in F}f.\Big\}
\end{equation}
In the following, we will use the fact $|\compset[r](0,\,0)|=1$.
\end{definition}

We proceed now to the explicit counting
\begin{prop}
Let us consider a recurrent configuration $r\in \recurnat$, with its
toppling domain $\domain[r]$, with $N:=|\domain[r]|$, and
associated sets  $\setconstraints[r]$ and $\compset[r](k,\,\ell)$. 
Then, for $i\in\llbracket 0,\delta_r\rrbracket$, the $\gam[r]i$ coefficient 
of \eqref{eq:gamma color} equals 
\begin{equation}
\label{eq:gamma}
\gam[r]i = \sum_{k=0}^{k_r^{\max}} (-1)^k \sum_{\ell=0}^i \binom{N-\ell}{i-\ell} \,
  \big|\compset[r](k,\,\ell)\big|
\end{equation}
where $k^{\max}_r$ is the largest value~$k$ such that $\compset[r](k,\,\ell)$ is not empty
for at least one value $\ell\leq i$. 
\end{prop}

\begin{proof}
The expression is obtained as the difference between the total number of colorings
and the number of non-legal colorings. 
The term $\binom{N}{i}$ ($k=0$) is the cardinal of $\coloring(r)$ and the following
terms count the non-legal colorings.
The difficulty lies in the final domains overlapping, and this can be solved using the 
\emph{inclusion-exclusion principle}.
Let us expose the application of this general idea to our problem.
We focus on a fixed $\gam[r]i$, so that all colorings $S\in\coloring(r)$ hereafter satisfy 
$\big|\{v\in\domain[r],\,S(v)=\instr{p}\}\big|= i$. 
We compute all non-legal colorings as follows
\begin{itemize}
\item Let us first choose $S$ non-legal, \textit{i.e.}
  there is a $f\in\setconstraints[r]$ where $S(v)=\instr{p}$. 
Given $f$, there are in total $\binom{N-|f|}{i-|f|}$ colorings naturally associated to $f$
and which are non-legal.
The degeneracy comes from the number of ways to assign the $i-|f|$ instructions $\instr{p}$ 
among the $N-|f|$, as $|f|$
instructions $\instr{p}$ have already been used on $f\subset\domain[r]$.
Summing over all $f\in\setconstraints[r]$ corresponds, in
\eqref{eq:gamma}, to the term $k=1$ and the sum over all final domain sizes $\ell\leq i$, 
weighted by their degeneracy $|\compset[r](k,l)|$ at $\ell$.
However, this term might include repetitions and so might overestimate the correction. 
In the second step, we balance this overshooting.
        
\item At the second step, we choose $S$ non-legal so that it contains $\instr{p}$
instructions over two different domains
$f_{1},f_{2}\in \setconstraints[r]$.
We can again associate to $V=f_{1}\cup f_{2}$ a number of $\binom{N-|V|}{i-|V|}$ non-legal colorings.
This balances the duplicate counts at the first step when we counted independently the natural degeneracy 
associated to $f_{1}$ and $f_{2}$.
Note that this imbalance occurs if and only if $|V|\leq i$, otherwise we already made a correct counting
at the previous step as we could not color both $f_1$ and $f_2$ with only $i$ instructions $\instr q$.
This compensating contribution should have an opposite sign to the first term, and corresponds 
in \eqref{eq:gamma} to term $k=2$ and all $\ell\leq i$. 
As one can realize, along this counting, we might perform over-counting again. 
Their could exist colorings $V$ including three different domains of $\setconstraints[r]$ 
and we shall proceed to another balancing step. 

\item We iterate this procedure for increasing $k$ until the balancing term equals $0$. 
At this maximal value $k^{\max}_r+1$, all union of $k^{\max}_r+1$ domains taken from $\setconstraints[r]$,
have a cardinal strictly greater than $i$ and are therefore not compatible with the constraint $i$,
which ends the computation of~$\gam[r]{i}$. 
\end{itemize}
\end{proof}

The solution~\eqref{eq:gamma} to the counting of paths is operational.
It is nonetheless still difficult to implement for 
large systems. We do not exclude that a more efficient way of counting is possible.
The amplitude of the terms in the sum~\eqref{eq:gamma} behaves non-monotonically, 
essentially because the degeneracy factor $|\compset[r](k,\,\ell)|$ tends to grow before decreasing.
The counting problem is notably hard due to the overlap of final domains.
We noticed that representatives of an
equivalence class $[r]$ generate a lot of overlapping and that this property scales, for fixed density parameters (\textit{e.g.} 
density of particles or stones) 
with the system size through both the toppling domain cardinal $|\domain[r]|$ 
and the number of constraints $|\setconstraints[r]|$ involved.
In the large size limit, however, one could rely on approximations to improve the efficiency
of the counting method. We are not aware, at the present time, of such approximations.

We conclude with a simple counting example for $L=3$ to illustrate the general method.
Using Figure \ref{fig:ex ct gamma} we compute $\compset[r]$ 
for $r=\zcfg{112}$ (cf Table \ref{tab:ex ct gamma}).

\begin{figure}[htbp]
\centering
\begin{tikzpicture}
\begin{scope}
\xaxis{2}{1,2,3}
\reprh{2,3,3}{0.5}{black}
\node[rectangle,draw] at (0.75,-0.75) {$\domain[r]$};
\end{scope}
\begin{scope}[xshift = 5cm]
\xaxis{2}{1,2,3}
\reprh{2,3,3}{0.5}{black}
\node[rectangle,draw] at (0.75,-0.75) {$\setconstraints[r]$};
\reprcstr[blue]{0/1,0/1,0/1}{0.2};
\reprcstr[green]{0/0,1/1,1/1}{0.2};
\reprcstr[red]{1/1,0/0,2/1}{0.2};
\reprcstr[blue!40]{1/1,2/1,0/0}{0.14};
\reprcstr[red!40]{0/0,2/1,2/1}{0.08};
\end{scope}
\end{tikzpicture}
\caption{Computing $\gam[r]i$ for $r:=\zcfg{112}$.
(Left) The domain $\domain[r]$ to color.
(Right) The set of constraints $\setconstraints[r]$ where each element $f$ has a unique color. 
There are $5$ different constraints (red, blue, green, rose and violet).}
\label{fig:ex ct gamma}
\end{figure}

\begin{table}[htbp]
    \centering
    \begin{tabular}{|c|c||c|c|c|c|c|c|}
        \hline
        \multicolumn{2}{|c}{\multirow{2}{*}{$\compset[r](k,\,\ell)$}} & \multicolumn{5}{|c|}{$\ell$}\\
        \hhline{~~-----}
         \multicolumn{2}{|c|}{}
          & 0 & 1 & 2 & 3 & 4 \\
        \hline
         \multirow{4}{*}{$k$}
          & 1 & 0 & 0 & 4 & 1 & 0 \\
        \hhline{~------}
          & 2 & 0 & 0 & 0 & 3 & 3 \\
        \hhline{~------}
          & 3 & 0 & 0 & 0 & 1 & 0 \\
        \hhline{~------}
          & 4 & 0 & 0 & 0 & 0 & 0\\
        \hline
    \end{tabular}
    \caption{$\compset[r](k,\,\ell)$ for $r=\zcfg{112}$ ($L=3$),
    where $\ell$ takes values up to $\delta_{r}=4$.}
    \label{tab:ex ct gamma}
\end{table}

Using Equations \eqref{eq:gamma} and calling $N=|\domain[r]|= 8$ we get 
\begin{align*}
    \gam[r]0 &= \binom{N}{0} = 1 \\
    \gam[r]1 &= \binom{N}{1} = 8 \\
    \gam[r]2 &= \binom{N}{2}-\bigg[4\binom{N-2}{0}\bigg] = 24 \\
    \gam[r]3 &= \binom{N}{3}-\bigg[4\binom{N-2}{1}+\binom{N-3}{0}\bigg]
            +\bigg[3\binom{N-3}{0}\bigg]-\bigg[\binom{N-3}{0}\bigg] = 33 \\
    \gam[r]4 &= \binom{N}{4}-\bigg[4\binom{N-2}{2}+\binom{N-3}{1}\bigg]
            +\bigg[3\binom{N-3}{1}+3\binom{N-4}{0}\bigg]\\
    &-\bigg[\binom{N-3}{1}\bigg]=18\\
\end{align*}
The coefficients coincide with the ones in Table \ref{tab:classes L3}. 
The latter have been derived using a straightforward evolution from $a_1\zcfg{2_L}$.
\section{Conclusion}

In this paper, we have revealed the complex structure of the stationary state of the Oslo model
under driven-dissipative condition.
To do so, we introduced a new representation of the configurations, the $g$-representation.
Our study benefits greatly from this representation
because it creates a clear and simple framework for reasoning.
We have identified static and dynamic invariants and an equivalence
relation among the recurrent configurations. The set of recurrent configurations, $\recur$, is organized into
equivalence classes. In each of these classes, the configurations are obtained after similar sequences of topplings
from the maximal configuration, such that their statistical weight are related by a simple expression. 
Each equivalence class has a natural configuration $\tilde r$. 
The number of natural configurations~$|\recurnat|=2^L$ is equal to the number of equivalence classes, whereas
the number of recurrent configurations is $|\recur|=F_{2L}$
the $(2L)^{\text{th}}$ term of the Fibonacci sequence. 
Since $F_{2L} \sim 2.618^{L}$, natural configurations form an exponentially small fraction of recurrent
configurations, which are themselves an exponentially small fraction of the $3^L$~stable configurations of the 
model.

Moreover, a recurrent configuration~$r\in\recur$ has a statistical 
weight $P_{r}$ in the stationary state
that factors into a polynomial in $p$ and $q=1-p$
\begin{equation*}
P_{r} = P_{\liftnat r}(p,\,q)  \, p^{\pi_r} q^{\theta_r}
\end{equation*}
where $\pi_r$ is the number of sites with slope~$z=2$ in $r$, $\theta_r$ is the number of topplings conserving the level in $g$-representation
when starting from $\liftnat r$, and $P_{\liftnat r}(p,\,q)$ is an homogeneous
polynomial constant over the equivalence class~$[\tilde r]$.

The coefficients of the polynomial~$P_{\liftnat r}$ count 
the number of stabilization sequences from the maximal 
configuration to any elements of the class~$[\tilde r]$.
These coefficients therefore describe the dynamics of the avalanches. 
We have mapped the computation of these coefficients to a coloring problem of the two-dimensional representation
of the dynamics, a technique inspired by recent progress in the study of Abelian sandpile models
\cite{hoffman_density_2024,forien_new_2025}.
Each avalanche between two fixed configurations corresponds to a binary coloring problem, and 
the necessary and sufficient conditions of existence of this coloring determine the existence of
an avalanche. 
Based on the coloring representation, we have computed the polynomials up to $L=5$, but the 
computation cost grows very rapidly with~$L$. 

In order to work on avalanche statistics for large~$L$,
we are investigating asymptotic equivalents of these coefficients and running numerical exact computations.
Our work based on the two-dimensional representation is so far restricted to the study of the
stationary state, defined as the result of the stabilization from the maximal configuration $a_1\zcfg{2_L}$. 
This approach can be extended to study the stable distribution from any recurrent configuration $a_1r$ 
with $r\in\recur$. As it was shown by Corral \cite{corral_calculation_2004}, this adds a difficulty
to the problem since, for a majority of configurations~$r\in\recur$,
only a small fraction of $\recur$ is accessible from a single avalanche starting with~$r$. 
We have developed an efficient algorithm reaching the exact stationary state
expression within few minutes of calculations on a single CPU for systems sizes less than~$18$ sites.
Our methods and techniques will be presented in a forthcoming article dedicated 
to the study the Oslo stationary state that was initiated in Ref.~\cite{corral_calculation_2004}
($L\leq 8$) and continued in \cite{pradhan_probability_2006,pradhan_sampling_2007} ($L\leq 12$).

Our work also suggests that more explicit and exact results could be derived together with numerical simulations.
In particular, in \cite{grassberger_oslo_2016}, hyperuniformity was probed in the $z$-representation of 
the system so that the probability measure of the stationary state must concentrate around specific 
configurations with close to periodic particle distributions.
Many configurations in $\recur$ should then be irrelevant for the stationary state problem in the large $L$ limit.
This could become the basis for asymptotic expressions of the probability measures.

\clearpage

\printbibliography[title={Bibliography}] 

@article{christensen_tracer_1996,
	title = {Tracer {Dispersion} in a {Self}-{Organized} {Critical} {System}},
	volume = {77},
	copyright = {http://link.aps.org/licenses/aps-default-license},
	issn = {0031-9007, 1079-7114},
	url = {https://link.aps.org/doi/10.1103/PhysRevLett.77.107},
	doi = {10.1103/PhysRevLett.77.107},
	language = {en},
	number = {1},
	urldate = {2025-01-07},
	journal = {Physical Review Letters},
	author = {Christensen, Kim and Corral, Álvaro and Frette, Vidar and Feder, Jens and Jøssang, Torstein},
	month = jul,
	year = {1996},
	pages = {107--110},
}

@article{frette_avalanche_1996,
	title = {Avalanche dynamics in a pile of rice},
	volume = {379},
	copyright = {http://www.springer.com/tdm},
	issn = {0028-0836, 1476-4687},
	url = {https://www.nature.com/articles/379049a0},
	doi = {10.1038/379049a0},
	language = {en},
	number = {6560},
	urldate = {2025-01-07},
	journal = {Nature},
	author = {Frette, Vidar and Christensen, Kim and Malthe-Sørenssen, Anders and Feder, Jens and Jøssang, Torstein and Meakin, Paul},
	month = jan,
	year = {1996},
	pages = {49--52}
}

@article{frette_sandpile_1993,
	title = {Sandpile models with dynamically varying critical slopes},
	volume = {70},
	copyright = {http://link.aps.org/licenses/aps-default-license},
	issn = {0031-9007},
	url = {https://link.aps.org/doi/10.1103/PhysRevLett.70.2762},
	doi = {10.1103/PhysRevLett.70.2762},
	language = {en},
	number = {18},
	urldate = {2025-01-07},
	journal = {Physical Review Letters},
	author = {Frette, Vidar},
	month = may,
	year = {1993},
	pages = {2762--2765}
}

@article{corral_calculation_2004,
	title = {Calculation of the transition matrix and of the occupation probabilities for the states of the {Oslo} sandpile model},
	volume = {69},
	copyright = {http://link.aps.org/licenses/aps-default-license},
	issn = {1539-3755, 1550-2376},
	url = {https://link.aps.org/doi/10.1103/PhysRevE.69.026107},
	doi = {10.1103/PhysRevE.69.026107},
	language = {en},
	number = {2},
	urldate = {2025-01-07},
	journal = {Physical Review E},
	author = {Corral, Álvaro},
	month = feb,
	year = {2004},
	pages = {026107}
}

@misc{chua_exact_2002,
	title = {Exact enumeration of the {Critical} {States} in the {Oslo} {Model}},
	url = {http://arxiv.org/abs/cond-mat/0203260},
	doi = {10.48550/arXiv.cond-mat/0203260},
	language = {en},
	urldate = {2025-01-07},
	publisher = {arXiv},
	author = {Chua, Alvin and Christensen, Kim},
	month = mar,
	year = {2002},
	note = {arXiv:cond-mat/0203260},
}

@article{dhar_steady_2004,
	title = {Steady {State} and {Relaxation} {Spectrum} of the {Oslo} {Rice}-pile},
	volume = {340},
	issn = {03784371},
	url = {http://arxiv.org/abs/cond-mat/0309490},
	doi = {10.1016/j.physa.2004.05.003},
	language = {en},
	number = {4},
	urldate = {2025-01-07},
	journal = {Physica A: Statistical Mechanics and its Applications},
	author = {Dhar, Deepak},
	month = sep,
	year = {2004},
	note = {arXiv:cond-mat/0309490},
	pages = {535--543},
}

@article{grassberger_oslo_2016,
	title = {Oslo model, hyperuniformity, and the quenched {Edwards}-{Wilkinson} model},
	volume = {94},
	copyright = {http://link.aps.org/licenses/aps-default-license},
	issn = {2470-0045, 2470-0053},
	url = {https://link.aps.org/doi/10.1103/PhysRevE.94.042314},
	doi = {10.1103/PhysRevE.94.042314},
	language = {en},
	number = {4},
	urldate = {2025-01-07},
	journal = {Physical Review E},
	author = {Grassberger, Peter and Dhar, Deepak and Mohanty, P. K.},
	month = oct,
	year = {2016},
	pages = {042314},
}

@article{bak_self-organized_1987,
	title = {Self-organized criticality: {An} explanation of the 1/ \textit{f} noise},
	volume = {59},
	copyright = {http://link.aps.org/licenses/aps-default-license},
	issn = {0031-9007},
	shorttitle = {Self-organized criticality},
	url = {https://link.aps.org/doi/10.1103/PhysRevLett.59.381},
	doi = {10.1103/PhysRevLett.59.381},
	language = {en},
	number = {4},
	urldate = {2025-01-07},
	journal = {Physical Review Letters},
	author = {Bak, Per and Tang, Chao and Wiesenfeld, Kurt},
	month = jul,
	year = {1987},
	pages = {381--384},
}

@misc{jarai_sandpile_2018,
	title = {Sandpile models},
	url = {http://arxiv.org/abs/1401.0354},
	doi = {10.48550/arXiv.1401.0354},
	language = {en},
	urldate = {2025-02-02},
	publisher = {arXiv},
	author = {Járai, Antal A.},
	month = sep,
	year = {2018},
	note = {arXiv:1401.0354 [math]},
}

@misc{hoffman_density_2024,
	title = {The density conjecture for activated random walk},
	url = {http://arxiv.org/abs/2406.01731},
	doi = {10.48550/arXiv.2406.01731},
	language = {en},
	urldate = {2025-03-06},
	publisher = {arXiv},
	author = {Hoffman, Christopher and Johnson, Tobias and Junge, Matthew},
	month = jul,
	year = {2024},
	note = {arXiv:2406.01731 [math]},
}

@misc{forien_new_2025,
	title = {A new proof of superadditivity and of the density conjecture for {Activated} {Random} {Walks} on the line},
	url = {http://arxiv.org/abs/2502.02579},
	doi = {10.48550/arXiv.2502.02579},
	language = {en},
	urldate = {2025-03-06},
	publisher = {arXiv},
	author = {Forien, Nicolas},
	month = feb,
	year = {2025},
	note = {arXiv:2502.02579 [math]},
}

@article{dhar_theoretical_2006,
	title = {Theoretical studies of self-organized criticality},
	volume = {369},
	copyright = {https://www.elsevier.com/tdm/userlicense/1.0/},
	issn = {03784371},
	url = {https://linkinghub.elsevier.com/retrieve/pii/S0378437106004006},
	doi = {10.1016/j.physa.2006.04.004},
	language = {en},
	number = {1},
	urldate = {2025-01-07},
	journal = {Physica A: Statistical Mechanics and its Applications},
	author = {Dhar, Deepak},
	month = sep,
	year = {2006},
	pages = {29--70},
}

@article{pruessner_oslo_2003,
	title = {Oslo rice pile model is a quenched {Edwards}-{Wilkinson} equation},
	volume = {67},
	copyright = {http://link.aps.org/licenses/aps-default-license},
	issn = {1063-651X, 1095-3787},
	url = {https://link.aps.org/doi/10.1103/PhysRevE.67.030301},
	doi = {10.1103/PhysRevE.67.030301},
	language = {en},
	number = {3},
	urldate = {2025-08-01},
	journal = {Physical Review E},
	author = {Pruessner, Gunnar},
	month = mar,
	year = {2003},
	pages = {030301}
}

@article{pruessner_exact_2004,
	title = {Exact solution of the totally asymmetric {Oslo} model},
	volume = {37},
	issn = {0305-4470, 1361-6447},
	url = {https://iopscience.iop.org/article/10.1088/0305-4470/37/30/005},
	doi = {10.1088/0305-4470/37/30/005},
	language = {en},
	number = {30},
	urldate = {2025-01-07},
	journal = {Journal of Physics A: Mathematical and General},
	author = {Pruessner, Gunnar},
	month = jul,
	year = {2004},
	pages = {7455--7471},
}

@article{dhar_exactly_1989,
	title = {Exactly solved model of self-organized critical phenomena},
	volume = {63},
	copyright = {http://link.aps.org/licenses/aps-default-license},
	issn = {0031-9007},
	url = {https://link.aps.org/doi/10.1103/PhysRevLett.63.1659},
	doi = {10.1103/PhysRevLett.63.1659},
	language = {en},
	number = {16},
	urldate = {2025-10-05},
	journal = {Physical Review Letters},
	author = {Dhar, Deepak and Ramaswamy, Ramakrishna},
	month = oct,
	year = {1989},
	pages = {1659--1662},
}

@article{pradhan_sampling_2007,
	title = {Sampling rare fluctuations of height in the {Oslo} ricepile model},
	volume = {40},
	issn = {1751-8113, 1751-8121},
	url = {http://arxiv.org/abs/cond-mat/0608144},
	doi = {10.1088/1751-8113/40/11/003},
	language = {en},
	number = {11},
	urldate = {2025-10-07},
	journal = {Journal of Physics A: Mathematical and Theoretical},
	author = {Pradhan, Punyabrata and Dhar, Deepak},
	month = mar,
	year = {2007},
	note = {arXiv:cond-mat/0608144},
	pages = {2639--2650},
}

@article{pradhan_probability_2006,
	title = {Probability distribution of residence times of grains in models of rice piles},
	volume = {73},
	copyright = {http://link.aps.org/licenses/aps-default-license},
	issn = {1539-3755, 1550-2376},
	url = {https://link.aps.org/doi/10.1103/PhysRevE.73.021303},
	doi = {10.1103/PhysRevE.73.021303},
	language = {en},
	number = {2},
	urldate = {2025-10-08},
	journal = {Physical Review E},
	author = {Pradhan, Punyabrata and Dhar, Deepak},
	month = feb,
	year = {2006},
	pages = {021303},
}

@article{aschwanden_25_2016,
	title = {25 {Years} of {Self}-{Organized} {Criticality}: {Solar} and {Astrophysics}},
	volume = {198},
	issn = {0038-6308, 1572-9672},
	shorttitle = {25 {Years} of {Self}-{Organized} {Criticality}},
	url = {http://link.springer.com/10.1007/s11214-014-0054-6},
	doi = {10.1007/s11214-014-0054-6},
	language = {en},
	number = {1-4},
	urldate = {2025-01-07},
	journal = {Space Science Reviews},
	author = {Aschwanden, Markus J. and Crosby, Norma B. and Dimitropoulou, Michaila and Georgoulis, Manolis K. and Hergarten, Stefan and McAteer, James and Milovanov, Alexander V. and Mineshige, Shin and Morales, Laura and Nishizuka, Naoto and Pruessner, Gunnar and Sanchez, Raul and Sharma, A. Surja and Strugarek, Antoine and Uritsky, Vadim},
	month = jan,
	year = {2016},
	pages = {47--166},
}

@misc{bouchaud_self-organized_2024,
	title = {The {Self}-{Organized} {Criticality} {Paradigm} in {Economics} \& {Finance}},
	url = {http://arxiv.org/abs/2407.10284},
	doi = {10.48550/arXiv.2407.10284},
	language = {en},
	urldate = {2025-01-12},
	publisher = {arXiv},
	author = {Bouchaud, Jean-Philippe},
	month = sep,
	year = {2024},
	note = {arXiv:2407.10284 [q-fin]},
}

@book{bak_how_1996,
	address = {New York, NY},
	title = {How {Nature} {Works}: {The} {Science} of {Self}-{Organized} {Criticality}},
	isbn = {978-1-4757-5426-1},
	shorttitle = {How {Nature} {Works}},
	language = {en},
	publisher = {Springer New York},
	author = {Bak, Per},
	year = {1996},
}
\end{document}